\documentclass[seceq]{ptptex}
\newcommand{\btem}{\bibitem}
\newcommand{\beq}{\begin{eqnarray}}
\newcommand{\eeq}{\end{eqnarray}}

\renewcommand{\d}{\partial}
\newcommand{\eps}{\epsilon}
\newcommand{\veps}{\varepsilon}

\newcommand{\bfW}{\mbox{{\boldmath $W$}}}
\newcommand{\bfX}{\mbox{{\boldmath $X$}}}

\newcommand{\bfF}{\mbox{{\boldmath $F$}}}

\def\Vec#1{\mbox{\boldmath$#1$}}
\def\Ren#1{\mbox{\boldmath$#1$}}

\title{
  First-principle Derivation of
  Stable First-order Generic-frame
  Relativistic Dissipative Hydrodynamic Equations from Kinetic Theory
  by Renormalization-group Method
}

\author{
  Kyosuke \textsc{Tsumura}${}^1$
  and
  Teiji \textsc{Kunihiro}${}^2$
}

\inst{
  ${}^1$Analysis Technology Center,
  Fujifilm Corporation,
  Kanagawa 250-0193, Japan
  \\
  ${}^2$Department of Physics,
  Kyoto University,
  Kyoto 606-8502, Japan
}

\abst{
  We derive 
  first-order relativistic dissipative
  hydrodynamic equations
  from relativistic Boltzmann equation
  on the basis of the renormalization-group (RG) method.
  We introduce a macroscopic-frame vector (MFV),
  which does not necessarily coincide with the flow velocity,
  to specify
  the local rest frame on which the macroscopic dynamics is described.
  The five hydrodynamic modes are naturally identified with the same number
  of the zero modes of the linearized collision operator, i.e.,
  the collision invariants.
  After defining the inner product in the function space spanned
  by the distribution function,
  the higher-order terms,
  which give rise to the dissipative effects,
  are constructed so that they are precisely orthogonal to the zero modes
  in terms of the inner product:
  Here,
  any ansatz's, such as the so-called conditions of fit
  used in the standard methods in an ad-hoc way, are not necessary.
  We elucidate that the Burnett term dose not affect
  the hydrodynamic equations owing to
  the very nature of the hydrodynamic modes as the zero modes.
  Then, applying the RG equation, we obtain the hydrodynamic equation
  in a generic frame specified by the MFV,
  as the coarse-grained and covariant equation.
  Our generic hydrodynamic equation
  reduces to hydrodynamic equations in various local rest frames,
  including the energy and particle frames
  with a choice of the MFV.
  We find that our equation in the energy frame
  coincides with that of Landau and Lifshitz,
  while the derived equation in the particle frame is slightly
  different from that of Eckart,
  owing to the presence of the dissipative internal energy.
  We prove that the Eckart equation can not be compatible
  with the underlying relativistic Boltzmann equation.
  The proof is made
  on the basis of the observation
  that the orthogonality condition to the zero modes
  coincides with the ansatz's
  posed on the dissipative parts of the energy-momentum tensor and the particle current
  in the phenomenological equations.
  We also present an analytic proof
  that all of our equations have a stable equilibrium state
  owing to the positive definiteness of the inner product.
}

\notypesetlogo

\begin{document}
  
  \maketitle
  
  \section{
    Introduction
  }
  \label{sec:1}
  Relativistic hydrodynamic equation is widely used
  in various fields of physics,
  especially in high-energy nuclear physics \cite{qcd001,qcd002,qcd003}
  and astrophysics \cite{ast001,ast002,ast003},
  and it seems that
  the study of the relativistic hydrodynamic equation
  with \textit{dissipative} effects
  is now becoming a central interest in these fields.

  For instance, it was shown that
  the dynamical evolution of
  the hot and/or dense QCD matter
  produced in the Relativistic Heavy Ion Collider (RHIC) experiments
  can be well described
  by the relativistic hydrodynamic simulations \cite{qcd002,qcd003}.
  The suggestion that the created matter may have only a tiny viscosity
  prompted an interest in the origin of the viscosity in the
  created matter to be described by the relativistic quantum field theory
  and also the dissipative hydrodynamic equations.
  Also the relativistic dissipative hydrodynamic equation
  has been applied to the various high-energy astrophysical phenomena,
  e.g.,
  the accelerated expansion of the universe by bulk viscosity of dark
  matter and/or dark energy \cite{ast002,ast003}.
  
  It is, however, noteworthy that
  we have not necessarily reached a full understanding of
  the theory of relativistic hydrodynamics for viscous fluids,
  although there have been many important studies
  since Eckart's pioneering work \cite{hen001}.
  
  \subsection{
    Fundamental problems with relativistic hydrodynamic equation
    for a viscous fluid
  }
  We may summarize the fundamental problems
  on the  relativistic dissipative hydrodynamic equations
  as follows:
  (a)~ambiguities and ad-hoc ansatz's in the definition of the flow velocity
  \cite{hen001,hen002,mic002,mic003,hyd001,mic005,muronga07,romatschke};
  (b)~unphysical instabilities of the equilibrium state \cite{hyd002};
  (c)~lack of causality \cite{mic005,mic001,hen003,mic004}.
  
  One might consider that
  the definition of the flow velocity should be merely
  a kind of the choice of the coordinate space
  to describe dynamics in a easy and practical way,
  and one can go back and forth
  between two different definitions by a Lorentz transformation.
  Furthermore,
  one might consider that
  the unphysical instabilities of the equilibrium state
  should be attributable to the lack of causality,
  and one can restore the instabilities
  automatically by solving the causality problem.
  Thus, although the relativistic dissipative hydrodynamic equation
  has been attracting a great interest,
  it seems that most works are concerned with
  the causality problem and examine
  phenomenological or semi-phenomenological causal equations with 
  some foundations.
  We argue, however, that
  the first two problems, (a) and (b),
  and the third one, (c), have different origins,
  and the first two must be resolved before the third one is addressed.
  The present paper is concerned with the first two problems,
  and the third one will be studied in the forthcoming paper \cite{next002}.
  
  (a)~The form of the relativistic dissipative hydrodynamic equation
  depends on the definition of the flow velocity,
  which is equivalent to the choice of the local rest frame of the fluid.
  Typical rest frames include the particle frame and the energy frame,
  and a phenomenological equation for the respective frame is constructed by
  Eckart \cite{hen001} and Landau and Lifshitz \cite{hen002},
  respectively.
  Here, the phenomenological construction is
  based on the following three ingredients;
  (A)~the particle-number and energy-momentum conservation laws,
  (B)~the law of the increase in entropy,
  and
  (C)~some specific assumptions on the choice of the flow.
  The first and second points are reasonable and
  used also in the construction of the non-relativistic Navier-Stokes equation,
  while the third point is specific for the relativistic case.
  To be more explicit,
  let $\delta T^{\mu\nu}$ and $\delta N^\mu$
  be the dissipative part of
  the energy-momentum tensor and the particle current,
  respectively.
  The point is that
  the forms of $\delta T^{\mu\nu}$ and $\delta N^\mu$
  are not determined uniquely
  only by
  the particle-number and energy-momentum conservation laws
  and
  the law of the increase in entropy
  without some physical ansatz's
  involving the flow velocity $u_{\mu}$
  with $u_{\mu}\,u^{\mu} = g^{\mu\nu} \, u_\mu \, u_\nu = 1$
  and $g^{\mu\nu} = \mathrm{diag}(+1,\,-1,\,-1,\,-1)$:
  The implicit assumptions made by Eckart \cite{hen001} are
  \begin{eqnarray}
    \label{eq:ansatz-i}
    \mathrm{(i)}  &&\,\,  \delta e \equiv u_\mu \, \delta T^{\mu\nu}
    \, u_\nu = 0,\\
    \label{eq:ansatz-ii}
    \mathrm{(ii)} &&\,\,  \delta n \equiv u_\mu \, \delta N^\mu = 0, \\
    \label{eq:ansatz-iii}
    \mathrm{(iii)}&&\,\, \nu_\mu \equiv \Delta_{\mu\nu} \, \delta N^\nu = 0,
  \end{eqnarray}
  where $\Delta_{\mu\nu} \equiv g_{\mu\nu} - u_{\mu} \, u_{\nu}$.
  The condition (i) claims the absence
  of the internal energy
  $\delta e$
  of the dissipative origin, while
  (ii) and (iii) require no dissipative particle-number density
  $\delta n$ and current $\nu^\mu$,
  respectively.
  On the other hand,
  those of Landau and Lifshitz \cite{hen002} consist of
  (i), (ii), and
  \begin{eqnarray}
    \label{eq:ansatz-iv}
    \mathrm{(iv)} &&\,\, Q_\mu \equiv \Delta_{\mu\nu}\,
    \delta T^{\nu\rho} \,u_\rho = 0.
  \end{eqnarray}
  In physical terms,
  (i) and (iv) claim the absence of the internal energy $\delta e$ and current $Q^{\mu}$ of
  dissipative origin.
  As one sees,
  the first two ansatz's are common for the two equations,
  and the third ones ((iii) and (iv)) are supposed to specify
  the respective local rest frame of the flow velocity.
  We note that the conditions
  (iii) and (iv) and hence the two of frames specified by these conditions
  can not be connected with each other by a Lorentz transformation.
  It is here noteworthy that
  there is a proposal by Stewart \cite{mic003}
  for the condition for the particle frame,
  as given by
  (ii), (iii), and
  \begin{eqnarray}
    \label{eq:ansatz-v}
    \mathrm{(v)} &&\,\, \delta T^{\mu}_{\,\,\,\mu} = \delta e -  3 \, \delta p = 0,
  \end{eqnarray}
  where $\delta p \equiv - \Delta_{\mu\nu}\,\delta T^{\mu\nu}/3$ is the dissipative pressure
  to be identified with the standard bulk pressure.
  Here, the condition (i) of Eckart is replaced by the different one (v),
  which claims a constraint between the dissipative internal energy $\delta e$ 
  and the dissipative pressure $\delta p$.
  One may ask if both the Eckart and Stewart ansatz's make sense or not.
  It is noteworthy that the most general derivation
  of the hydrodynamic equation on the basis of the phenomenological argument
  gives a class of equations which can allow the existence of the
  dissipative internal energy $\delta e$
  and the dissipative particle-number density $\delta n$ as well as
  the standard dissipative pressure $\delta p$,
  as shown by the present authors \cite{Tsumura:2009vm};
  a brief recapitulation of Ref.\citen{Tsumura:2009vm}
  is given in Appendix A.
  
  (b)~The unphysical instabilities of the equilibrium state
  might be attributed to the lack of causality,
  and Israel-Stewart's formalism
  is presently being examined in connection to this problem
  \cite{mic004,muronga07,chaudhuri,romatschke}.
  Although their equation may get rid of the instability problem
  with a choice of the relaxation times,
  as shown in Ref.\citen{hyd002},
  we emphasize that
  there exists no connection between
  the unphysical instabilities and the lack of causality.
  In fact,
  the Landau-Lifshitz equation is free from the instabilities of the equilibrium state
  in contrast of the Eckart equation.
  Furthermore, one should notice that
  the causal equation by Israel and Stewart
  is an extended version of the Eckart equation
  and hence it can naturally exhibit unphysical instabilities
  depending on the values of
  transport coefficients and relaxation times contained
  in the equation \cite{biro07}.
  
  \subsection{
    Strategy of the present work:
    separation of the scales of dynamics
    and construction of hydrodynamic equation
    as the infrared effective dynamics
  }
  Here, we note
  the hierarchy of the dynamics of the time evolution
  of a many-body system:
  In the beginning of the time evolution of a prepared state,
  the whole dynamical evolution of the system
  will be governed by Hamiltonian dynamics
  that is time-reversal invariant.
  When the system becomes old,
  the dynamics is relaxed to the kinetic regime,
  where the time-evolution system is well described by
  a truncation of the BBGKY
  (Bogoliubov-Born-Green-Kirkwood-Yvon) hierarchy \cite{bbgky};
  the Boltzmann equation composed of one-body distribution function
  describes a coarse grained slower dynamics,
  in which time-reversal invariance is lost.
  Then, as the system is further relaxed,
  the time evolution will be described
  in terms of the hydrodynamic quantities,
  i.e.,
  the flow velocity,
  the particle-number density,
  and the local temperature.
  In this sense,
  the hydrodynamics is the infrared asymptotic dynamics of
  the kinetic equation.

  Now the relativistic Boltzmann equation is a typical kinetic equation,
  which
  is manifestly Lorentz invariant
  and free from the instability and causal problems \cite{mic001}.
  Thus, one sees that a natural way to resolve
  the ambiguities in the definition of the flow velocity
  and the unphysical instabilities of the equilibrium state
  is to derive the relativistic dissipative hydrodynamic equation
  from an underlying relativistic kinetic equation.
  Indeed,
  there have been some vigorous attempts
  to derive the phenomenological equations
  from the relativistic Boltzmann equation;
  for instance,
  with use of the Chapman-Enskog expansion method \cite{chapman}
  and the Maxwell-Grad moment method \cite{grad}.
  We would say, however, that
  these works in the microscopic approaches are not fully satisfactory:
  Although the past works certainly succeeded
  in identifying the assumptions and/or approximations
  to reproduce the known hydrodynamic equations
  by Eckart, Landau and Lifshitz, Stewart, and Israel,
  the physical meaning and foundation of these
  assumptions/approximations remain obscure,
  and thus the uniqueness of those hydrodynamic equations
  has never been elucidated as
  the long-wavelength and low-frequency limit of the underlying dynamics.
  Indeed, the standard derivation of relativistic hydrodynamic equations
  based on the Chapman-Enskog expansion or Maxwell-Grad moment method \cite{mic001}
  utilizes the ansatz's given
  by (i) $\sim$ (v) as the constraints
  on the distribution function as the solution of
  the relativistic Boltzmann equation,
  rather than consequences of the derivation.
  Their validity or the fundamental compatibility with the underlying Boltzmann equation
  has never been questioned nor addressed.
  This unsatisfactory situation rather reveals
  the incompleteness
  of the Chapman-Enskog expansion method and the Maxwell-Grad moment methods
  themselves as a reduction theory of the dynamics.
  
  Thus, the form of relativistic dissipative hydrodynamic equations
  is still controversial and far from being established.
  The origins of the difficulty of the derivation are identified as
  the absence of an appropriate coarse-graining method
  that keeps the Lorentz covariance
  and is applicable to the relativistic Boltzmann equation.
  It should be mentioned here that van Kampen \cite{mic005}
  applied his reduction theory to derive
  a relativistic hydrodynamic equation.
  The resultant equation was, unfortunately,
  of a noncovariant form.

  In this work,
  we try to derive the relativistic dissipative hydrodynamic equation
  from the relativistic Boltzmann equation
  in a more natural and systematic way:
  For that, it is essential to adopt
  a powerful reduction theory of the dynamics \cite{kuramoto}.
  As such a reduction theory,
  we take the ``renormalization-group (RG) method''.
  In Ref.'s \citen{rgm001,env001,env002,env005,qm,env006,env007,env008},
  it has been shown that
  the RG method is a powerful resummation method
  and also gives a systematic reduction theory of the dynamics
  leading to the coarse graining of temporal and spatial scales,
  which are the key concepts in the construction
  of infrared effective theories
  \footnote{
    A brief account of the RG method is given in Appendix B
    using an example for self-containedness.
  }.
  Indeed, the RG method is already applied satisfactorily
  to the derivation of the Navier-Stokes equation
  from the (non-relativistic) Boltzmann equation,
  with no heuristic assumption \cite{env007,env008}.
  The RG method thus should be most suitable for the present purpose
  to derive covariant relativistic dissipative hydrodynamic equations.
  It is also expected that the physical meanings and the validity
  of the ansatz's posed in the phenomenological derivation will
  be elucidated in the process of the reduction in the RG method.
  
  In Ref.\citen{env009},
  the present authors and K. Ohnishi applied the RG method
  to derive the relativistic dissipative hydrodynamic equations
  for the first time:
  A macroscopic-frame vector was introduced
  with which the derivation of a coarse-grained covariant equation
  is made possible,
  and thus
  the so-called \textit{first-order} relativistic dissipative hydrodynamic
  equations were successfully derived,
  in which
  the dissipative effects are taken into account up to the first order.
  The five hydrodynamic variables naturally
  correspond to the zero modes of the linearized collision operator.
  The deviations from the local equilibrium
  are given by the functions that are precisely orthogonal to the zero modes
  with an inner product for the distribution functions.
  It was found that
  the various local rest frames of the flow velocity can be realized by
  a choice of the macroscopic-frame vector.
  Subsequently,
  the present authors \cite{env010} showed that
  the derived hydrodynamic equation in the particle frame
  is different from the Eckart and Stewart equations,
  and has the stable equilibrium state
  in a numerical calculation
  for a rarefied gas where dissipative effects are most significant.
  
  \subsection{
    Purpose of the present paper
  }
  The purpose of this paper is as follows:
  (1)~We present a detailed and full account of the derivation of
  the first-order relativistic dissipative hydrodynamic equations
  in generic frames
  from the relativistic Boltzmann equation
  on the basis of the RG method.
  Moreover,
  (2)~we elaborate and revise some parts of the derivation
  and thereby make it more transparent:
  We clarify the essential importance
  to properly define the inner product of the distribution functions,
  in particular to make it positive-definite,
  which was not recognized in the previous presentation.
  We also show using simple properties of the inner product
  that the usually problematic Burnett term
  does not contribute to the hydrodynamic equations
  on account of
  the fact that the hydrodynamic modes
  are described by the zero modes of the linearized collision operator.
  (3)~We show that the ansatz's
  (i) $\sim$ (v) for the dissipative currents
  exactly correspond to the orthogonality conditions
  of the perturbed distribution function to
  the zeroth-order solution,
  and that the Eckart constraint (i) $\sim$ (iii)
  can not be compatible with the underlying Boltzmann equation as a corollary.
  (4)~We also fully examine the properties
  and its advantageous nature
  of the derived equation in a generic frame.
  We give a general proof without recourse to numerical calculations
  that the hydrodynamic equations obtained in our formalism
  have the stable equilibrium state even in the particle (Eckart) frame
  on the basis of the positive-definiteness of the inner product.
  
  This paper is organized as follows:
  In \S\ref{sec:preliminaries},
  after a brief account of the basic ingredients of
  the relativistic Boltzmann equation,
  we introduce the macroscopic-frame vector,
  and
  summarize the ad-hoc aspects in the standard methods such as
  the Chapman-Enskog expansion and Maxwell-Grad moment methods.
  In \S\ref{sec:2},
  with use of the RG method,
  we reduce
  the relativistic Boltzmann equation
  to a generic form of
  the relativistic dissipative hydrodynamic equation,
  whose frame is not specified.
  In \S\ref{sec:3},
  we show that
  the obtained equation
  reduces to
  the relativistic dissipative hydrodynamic
  equations in various frames
  with a choice of the macroscopic-frame vector,
  including the particle one and the energy one.
  Then, we compare our equations with those
  proposed by Eckart, Landau-Lifshitz, and Stewart.
  In \S\ref{sec:4} and \S\ref{sec:5},
  we examine some properties of our equations,
  concerning transport coefficients and frames.
  In \S\ref{sec:6},
  we present a proof that
  all of our equations have the stable equilibrium state.
  The last section is devoted to a summary and concluding remarks.
  In Appendix A,
  we show that the most general form of the hydrodynamic
  equation as derived in the standard phenomenological way may admit
  the dissipative internal energy, pressure, and particle-number density.
  In Appendix B, after giving a brief account of the RG method
  using an example,
  we present a general ground of this method.
  In Appendix C,
  we present a detailed derivation of the first-excited modes
  in a generic local rest frame.
  
  \setcounter{equation}{0}
  \section{
    Preliminaries
  }
  \label{sec:preliminaries}
  After a brief account of the basic properties of the relativistic Boltzmann equation,
  we introduce a macroscopic-frame vector
  to specify a local rest frame
  on which the macroscopic dynamics is described:
  We shall see that the introduction of the macroscopic-frame vector
  enables us to have a coarse-grained and covariant equation.
  As in the standard method like the Chapman-Enskog expansion method and others \cite{mic001},
  we take the spatial inhomogeneity as the origin of the dissipation.
  We shall clarify the ad-hoc aspects of the ansatz's made
  in the standard Chapman-Enskog expansion and Maxwell-Grad moment methods
  for the derivation of the relativistic hydrodynamic equations.
  
  \subsection{
    Relativistic Boltzmann equation
  }
  As a simple example,
  we shall treat the classical relativistic system composed of identical particles
  \footnote{
    An extension to multi-component systems is possible
    and will be presented elsewhere.
  }.
  Then, the relativistic Boltzmann equation \cite{mic001} for such a system reads
  \begin{eqnarray}
    \label{eq:1-001}
    p^\mu \, \partial_\mu f_p(x) = C[f]_p(x),
  \end{eqnarray}
  where $f_p(x)$ denotes the one-particle distribution function
  defined in the phase space $(x \,,\, p)$
  with
  $x^\mu$ being the space-time coordinate and
  $p^\mu$ being the four momentum of the on-shell particle;
  $p^\mu \, p_\mu = p^2 = m^2$ and $p^0 > 0$.
  The right-hand side of Eq.(\ref{eq:1-001}) is called the collision integral,
  \begin{eqnarray}
    \label{eq:1-002}
    C[f]_p(x) \equiv \frac{1}{2!} \, \sum_{p_1} \, \frac{1}{p_1^0} \,
    \sum_{p_2} \, \frac{1}{p_2^0} \, \sum_{p_3} \, \frac{1}{p_3^0} \, 
    \omega(p \,,\, p_1|p_2 \,,\, p_3)
    \Big( f_{p_2}(x) \, f_{p_3}(x) - f_p(x) \, f_{p_1}(x) \Big),\nonumber\\
  \end{eqnarray}
  where $\omega(p \,,\, p_1|p_2 \,,\, p_3)$ denotes
  the transition probability due to the microscopic two-particle interaction.
  To make explicit the correspondence to the general formulation given in Ref.\citen{env006},
  we treat the momentum as a discrete variable,
  but the summation with respect to the momentum
  is interpreted as the integration;
  $\sum_{q} \equiv \int \!\! \mathrm{d}^3\Vec{q}$
  with $\Vec{q}$ being the spatial components of the four momentum $q^\mu$.
  We remark that
  the transition probability $\omega(p \,,\, p_1|p_2 \,,\, p_3)$ contains
  the delta functions representing the energy-momentum conservation,
  \begin{eqnarray}
    \label{eq:1-003}
    \omega(p \,,\, p_1|p_2 \,,\, p_3) \propto \delta^4(p + p_1 - p_2 - p_3),
  \end{eqnarray}
  and also has the symmetric properties
  due to the indistinguishability of the particles
  and the time reversal invariance of
   the microscopic transition probability,
  \begin{eqnarray}
    \label{eq:1-004}
    \omega(p \,,\, p_1|p_2 \,,\, p_3) = \omega(p_2 \,,\, p_3|p \,,\, p_1)
    = \omega(p_1 \,,\, p|p_3 \,,\, p_2) = \omega(p_3 \,,\, p_2|p_1 \,,\, p).
  \end{eqnarray}
  It should be stressed here that
  we have confined ourselves to the case
  in which the particle number is conserved in the collision process.
  
  The property of the transition probability shown in Eq.(\ref{eq:1-004})
  leads to the following identity satisfied for an arbitrary vector $\varphi_p(x)$,
  \begin{eqnarray}
    \label{eq:coll-symm}
    \sum_p \, \frac{1}{p^0} \, \varphi_p(x) \, C[f]_p(x)
    &=& \frac{1}{2!} \,\frac{1}{4}\, \sum_{p} \, \frac{1}{p^0}\,
    \sum_{p_1} \, \frac{1}{p^0_1}\,
    \sum_{p_2} \, \frac{1}{p^0_2}\,
    \sum_{p_3} \, \frac{1}{p^0_3}\,
    \nonumber\\
    &&
    \times
    \omega(p \,,\, p_1|p_2 \,,\, p_3)\, 
    \Big(\varphi_p(x) \,+\,\varphi_{p_1}(x)\,- \,\varphi_{p_2}(x)\, -\,\varphi_{p_3}(x)\Big)
    \nonumber \\
    &&
    \times
    \Big( f_{p_2}(x) \, f_{p_3}(x) - f_p(x) \, f_{p_1}(x) \Big).
  \end{eqnarray}
  
  A function $\varphi_p(x)$ is called a collision invariant
  (or summational invariant \cite{mic001})
  when it satisfies the following equation
  \begin{eqnarray}
    \label{eq:coll-inv}
    \sum_p \, \frac{1}{p^0} \, \varphi_p(x) \, C[f]_p(x) = 0.
  \end{eqnarray}
  As is easily confirmed by using the formula (\ref{eq:coll-symm})
  and the property (\ref{eq:1-003}),
  $\varphi_p = 1$ and $p^{\mu}$ are collision invariants;
  \begin{eqnarray}
    \label{eq:1-005}
    \sum_p \, \frac{1}{p^0} \, C[f]_p(x) &=& 0,\\
    \label{eq:1-006}
    \sum_p \, \frac{1}{p^0} \, p^\mu \, C[f]_p(x) &=& 0,
  \end{eqnarray}
  which represent, of course, the conservation of
  the particle number, energy, and momentum
  by the collision process, respectively.
  We also see that the linear combination of these collision invariants as given by 
  \begin{eqnarray}
    \varphi_p(x) = a(x) + p^{\mu} \, b_{\mu}(x),
  \end{eqnarray}
  is also a collision invariant 
  with $a(x)$ and $b_{\mu}(x)$ being arbitrary functions of $x$.
  This form is, in fact, known to be the most general form of a collision invariant \cite{mic001}.
  
  On account of Eq.'s (\ref{eq:1-005}) and (\ref{eq:1-006}),
  we have the continuity or balance equations
  for the particle current $N^\mu$
  and the energy-momentum tensor $T^{\mu\nu}$,
  \begin{eqnarray}
    \label{eq:1-007}
    \partial_\mu N^\mu(x)
    \equiv
    \partial_\mu \Bigg[\sum_p \, \frac{1}{p^0} \, p^\mu \, f_p(x)\Bigg]
    &=& 0,\\
    \label{eq:1-008}
    \partial_\nu T^{\mu\nu}(x)
    \equiv
    \partial_\nu \Bigg[\sum_p \, \frac{1}{p^0} \, p^\mu \, p^\nu \, f_p(x)\Bigg]
    &=& 0,
  \end{eqnarray}
  respectively.
  It is noted that
  while these equations have the same forms as the hydrodynamic equations,
  nothing about the dynamical properties is contained in these equations
  before the evolution of the distribution function $f_p(x)$
  is obtained by solving Eq.(\ref{eq:1-001}).
  In the standard Chapman-Enskog expansion method \cite{mic001},
  these balance equations are rather used
  to obtain the time derivatives
  of the distribution function written in terms of the hydrodynamic quantities,
  order by order.
  
  The entropy current is defined by
  \begin{eqnarray}
    \label{eq:1-009}
    S^\mu(x) \equiv - \sum_p \, \frac{1}{p^0} \, p^\mu \, f_p(x) \, (\ln f_p(x) - 1).
  \end{eqnarray}
  Using the relativistic Boltzmann equation (\ref{eq:1-001}),
  the divergence of the entropy current reads
  \begin{eqnarray}
    \label{eq:1-010}
    \partial_\mu S^\mu(x) =  - \sum_p \, \frac{1}{p^0} \, C[f]_p(x) \, \ln f_p(x).
  \end{eqnarray}
  The above equation tells us that
  $S^\mu$ is conserved only if $\ln f_p(x)$ is a collision invariant,
  or a linear combination of the basic collision invariants $(1 \,,\, p^\mu)$
  as
  \begin{eqnarray}
    \label{eq:1-011}
    \ln f_p(x) = \alpha(x) + p^\mu \, \beta_\mu(x),
  \end{eqnarray}
  with $\alpha(x)$ and $\beta_\mu(x)$ being arbitrary functions of $x$.
  In other words,
  the entropy-conserving distribution function is parametrized as
  \begin{eqnarray}
    \label{eq:Juettner}
    f_p(x)
    = \frac{1}{(2\pi)^{3}} \,
    \exp \Bigg[ \frac{\mu(x)
        - p^\mu \, u_\mu(x)}{T(x)} \Bigg]
    \equiv f^{\mathrm{eq}}_p(x),
  \end{eqnarray}
  with $u^\mu(x) \, u_\mu(x) = 1$,
  which is identified with the local equilibrium distribution function
  called the Juettner function \cite{juettner}
  (the relativistic analog of the Maxwellian):
  $T(x)$, $\mu(x)$, and $u^\mu(x)$ in Eq.(\ref{eq:Juettner})
  should be interpreted as the local temperature,
  the chemical potential,
  and the flow velocity, respectively.
  
  We note that for the local equilibrium distribution $f^{\mathrm{eq}}_p(x)$ 
  the collision integral identically vanishes,
  \begin{eqnarray}
    \label{eq:1-012}
    C[f^{\mathrm{eq}}]_p(x) = 0,
  \end{eqnarray}
  due to the energy-momentum conservation
  implemented in the transition probability (\ref{eq:1-003}).
  
  \subsection{
    Introduction of macroscopic-frame vector
  }
  We are now in a position to introduce
  the key ingredient in the present work.
  
  Since we are interested in the hydrodynamic regime
  where the time and space dependence of the physical quantities are small,
  we try to solve Eq.(\ref{eq:1-001})
  in the hydrodynamic regime where the space-time variation of $f_p(x)$ is small
  and the space-time scales are coarse-grained from those in the kinetic regime.
  To make a coarse graining with the Lorentz covariance being retained,
  we introduce a 
  time-like Lorentz vector denoted by
  \begin{eqnarray}
    \label{eq:1-013}
    \Ren{a}^\mu,
  \end{eqnarray}
  with
  \begin{eqnarray}
    \label{eq:1-013-2}
    \Ren{a}^0\,>\,0.
  \end{eqnarray}
  Thus, $\Ren{a}^\mu$
  specify the covariant but macroscopic coordinate system where 
  the velocity field of the hydrodynamic flow is defined:
  Since such a coordinate system is called \textit{frame},
  we call $\Ren{a}^\mu$ the \textit{macroscopic-frame vector}. 
  Although $\Ren{a}^\mu$ may depend on the momentum $p$ and the
  space-time coordinate $x$, i.e.,
  \begin{eqnarray}
    \label{eq:1-014}
    \Ren{a}^\mu = \Ren{a}^\mu_p(x),
  \end{eqnarray}
  the time variation of it is supposed to be much smaller
  than that of the microscopic processes.
  We shall see that
  the separation of the scales between the kinetic and hydrodynamic regimes
  can be nicely achieved  by the RG method.
  It should be stressed here that although a macroscopic vector is introduced 
  also in the standard Chapman-Enskog expansion method \cite{mic001},
  the vector is identified as the flow velocity $u^{\mu}(x)$ from the outset.
  In our case,
  $\Ren{a}^\mu_p(x)$ does not necessarily coincide with
  the flow velocity $u^{\mu}(x)$.

  Keeping in mind the above scale difference,
  we decompose the derivative $\partial^{\mu}$ into time-like and space-like ones
  in terms of the macroscopic-frame vector.
  Defining a projection operator to the space-like vector by
  \begin{eqnarray}
    \Ren{\Delta}^{\mu\nu}_p(x)
    \equiv  g^{\mu\nu} - \frac{\Ren{a}_p^\mu(x) \, \Ren{a}_p^\nu(x)}{\Ren{a}_p^2(x)},
  \end{eqnarray}
  we have
  \begin{eqnarray}
    \label{eq:1-015}
    \partial^\mu = \frac{\Ren{a}^\mu_p(x)\,\Ren{a}^\nu_p(x)}{\Ren{a}^2_p(x)}\,
    \partial_\nu
    +\Ren{\Delta}^{\mu\nu}_p(x)\,\partial_\nu
    =  \Ren{a}_p^{\mu}(x)\frac{\partial}{\partial \tau} + \frac{\partial}{\partial \sigma_{\mu}},
  \end{eqnarray}
  with
  \begin{eqnarray}
    \label{eq:1-017}
    \frac{\partial}{\partial\tau} &\equiv& \frac{1}{\Ren{a}^2_p(x)} \,
    \Ren{a}^\nu_p(x)\,\partial_\nu  ,\\
    \label{eq:1-018}
    \frac{\d}{\d \sigma_{\mu}} &\equiv&
    \Ren{\Delta}^{\mu\nu}_p(x)\,\partial_\nu \equiv \Ren{\nabla}^\mu.
  \end{eqnarray}
  
  Then, the relativistic Boltzmann equation (\ref{eq:1-001}) 
  in the new coordinate system $(\tau \,,\, \sigma^\mu)$
  is written as
  \begin{eqnarray}
    \label{boltzmanneq}
    p \cdot \Ren{a}_p(\tau \,,\, \sigma) \, \frac{\partial}{\partial \tau} f_p(\tau \,,\, \sigma)
    + p \cdot \Ren{\nabla} f_p(\tau \,,\, \sigma)
    = C[f]_p(\tau \,,\, \sigma),
  \end{eqnarray}
  where $\Ren{a}_p^\mu(\tau \,,\, \sigma) \equiv \Ren{a}_p^\mu(x)$
  and $f_p(\tau \,,\, \sigma) \equiv f_p(x)$.
  We remark the prefactor of the time derivative
  is a Lorentz scalar and positive definite;
  \begin{eqnarray}
    \label{eq:p-a-positive}
    p \cdot \Ren{a}_p(\tau \,,\, \sigma) > 0,
  \end{eqnarray}
  which is easily verified by taking the rest frame of $p^0$.
  
  Since we are interested in a hydrodynamic solution
  to Eq.(\ref{boltzmanneq}) as mentioned above,
  we shall convert Eq.(\ref{boltzmanneq}) into
  \begin{eqnarray}
    \label{eq:1-019}
    \frac{\partial}{\partial \tau} f_p(\tau \,,\, \sigma)
    = \frac{1}{p \cdot \Ren{a}_p(\tau \,,\, \sigma)} \, C[f]_p(\tau \,,\, \sigma)
    - \varepsilon \, \frac{1}{p \cdot \Ren{a}_p(\tau \,,\, \sigma)}
    \, p \cdot \Ren{\nabla} f_p(\tau \,,\, \sigma),
  \end{eqnarray}
  where a small quantity $\varepsilon$ has been introduced to express
  that the space derivatives are small for the system which we are
  interested in;
  $\varepsilon$ is called the non-uniformity parameter \cite{mic001}
  and may be identified with
  the ratio of the average particle distance over the mean free path,
  i.e., the Knudsen number.
  
  Since $\varepsilon$ appears
  in front of the second term of the right-hand side of Eq.(\ref{eq:1-019}),
  the relativistic Boltzmann equation
  has a form to which the perturbative expansion is applicable.
  In fact,
  this form of the Boltzmann equation but with $\Ren{a}^{\mu}_p = u^{\mu}$
  is also the starting point for
  the standard Chapman-Enskog expansion method \cite{mic001},
  together with an ad-hoc ansatz on the order of the time derivative of the distribution function.
  We stress that this seemingly mere rewrite of the equation has a physical significance;
  it expresses a natural assumption
  that only the spatial inhomogeneity over distances of the order of the mean free path
  is the origin of the dissipation.
  We remark that
  the RG method applied to non-relativistic Boltzmann equation
  with the corresponding assumption leads 
  to the Navier-Stokes equation \cite{chapman,kuramoto,env007,env008};
  the present approach is simply a covariantization of the non-relativistic case.
  
  For setting up the perturbative expansion
  in a consistent way with the physical picture of the origin of the dissipation,
  we shall take the coordinate system
  where $\Ren{a}_p^\mu(\tau \,,\, \sigma)$ has no $\tau$ dependence, i.e.,
  \begin{eqnarray}
    \Ren{a}_p^\mu(\tau \,,\, \sigma)= \Ren{a}_p^\mu(\sigma).
  \end{eqnarray}
  Then, Eq.(\ref{eq:1-019}) takes the following form,
  \begin{eqnarray}
    \label{eq:start}
    \frac{\partial}{\partial \tau} f_p(\tau \,,\, \sigma)
    = \frac{1}{p \cdot \Ren{a}_p(\sigma)} \, C[f]_p(\tau \,,\, \sigma)
    - \varepsilon \, \frac{1}{p \cdot \Ren{a}_p(\sigma)}
    \, p \cdot \Ren{\nabla} f_p(\tau \,,\, \sigma).
  \end{eqnarray}
  
  Now let $q_p(\sigma)$
  is a physical quantity of a particle with a momentum $p^\mu$ at a position $\sigma^{\mu}$;
  then the total amount of the quantity is given by
  \begin{eqnarray}
    Q_p(\tau) = \int\!\! \mathrm{d}^3\sigma \, q_p(\sigma) \, f_p(\tau\,,\,\sigma).
  \end{eqnarray}
  When $q_p(\sigma) = p \cdot \Ren{a}_p(\sigma)$,
  the time variation of $Q_p(\tau)$ is given by
  \begin{eqnarray}
    \label{eq:transported}
    \frac{\mathrm{d}}{\mathrm{d}\tau}Q_p(\tau)&=&
    \int \!\! \mathrm{d}^3\sigma\, p\cdot \Ren{a}_p(\sigma) \,
    \frac{\d}{\d\tau}f_p(\tau\, ,\,\sigma),\nonumber \\
    &=& - \varepsilon \, p^\mu \, \int \!\! \mathrm{d}^3\sigma\,
    \frac{\partial}{\partial\sigma^\mu} f_p(\tau \,,\, \sigma)
    +\int \!\! \mathrm{d}^3\sigma\, C[f]_p(\tau\,,\,\sigma)
    ,\nonumber \\
    &=& \int \!\! \mathrm{d}^3\sigma \, C[f]_p(\tau\,,\,\sigma),
  \end{eqnarray}
  where we have used Eq.(\ref{eq:start})
  and neglected the contribution from the spatial boundary of the system.
  Here, we consider a trajectory of one particle
  in the interval between each collisions:
  Substituting $C[f]_p(\tau\,,\,\sigma) = 0$ into Eq.(\ref{eq:transported}),
  we find that
  $Q_p(\tau)$ is a conserved quantity and $q_p(\sigma)$ is a
  corresponding density.
  Thus,
  the form of the relativistic Boltzmann equation as given in Eq.(\ref{eq:start})
  tells us that the physical quantity that is transported by each particle in the system
  is given by
  \begin{eqnarray}
    \label{eq:quantity}
    p \cdot \Ren{a}_p(\sigma).
  \end{eqnarray}
  
  It is to be noted that
  we can control what is the flow represented in our theory
  by varying the specific expression of $\Ren{a}_p^\mu(\sigma)$.
  Because of this freedom inherent in our coordinate system,
  our theory may lead to various hydrodynamic equations,
  including the ones in the energy and particle frames for non-ideal
  fluids.
  
  \subsection{
    A brief description of the standard methods
  }
  Before developing our analysis based on the RG method,
  we briefly summarize the ad-hoc aspects in the standard methods
  such as the Chapman-Enskog expansion and Maxwell-Grad moment methods.
  
  In the standard Chapman-Enskog expansion method,
  one starts from the following form of the Boltzmann equation,
  \begin{eqnarray}
    p\cdot u\, D_{\mathrm{CE}} f_p(x) = - \veps \, p \cdot \nabla_{\mathrm{CE}} f_p(x) + C[f]_p(x),
  \end{eqnarray}
  where
  $D_{\mathrm{CE}} \equiv u^\mu \, \partial_\mu$
  and
  $\nabla^\mu_{\mathrm{CE}} \equiv \Delta_{\mathrm{CE}}^{\mu\nu} \, \partial_{\nu}$
  with $\Delta_{\mathrm{CE}}^{\mu\nu} \equiv g^{\mu\nu} - u^{\mu} \, u^{\nu}$.
  Then, one makes the perturbative expansion
  \begin{eqnarray}
    f_p(x) &=& f^{(0)}_p(x) + \veps \, f^{(1)}_p(x) + \veps^2 \, f^{(2)}_p(x) + \cdots 
    \equiv f^{(0)}_p(x) + \delta f_p(x)  ,\\
    D_{\mathrm{CE}}f_p(x) &=& \veps \, D_{\mathrm{CE}} f_p^{(1)}(x)
    + \veps^2 \, D_{\mathrm{CE}}f_p^{(2)}(x) + \cdots.
  \end{eqnarray}
  Then, it is found that the zeroth-order solution is given by the local equilibrium 
  distribution function, i.e., the Juettner function given by Eq.(\ref{eq:Juettner});
  \begin{eqnarray}
    f^{(0)}_p(x) = f^{\mathrm{eq}}_p(x).
  \end{eqnarray}
  
  It is customary \cite{mic001},
  without any physical foundation,
  to assume that the particle-number density and internal energy
  in the non-equilibrium state is the same as those
  in the local equilibrium state, and set as follows;
  \begin{eqnarray}
    \label{eq:cond-fit-n-1}
    n &\equiv& u_{\mu} \, \Bigg[
      \sum_p \, \frac{1}{p^0} \,  p^{\mu} \, f_p
      \Bigg]
    =
    u_{\mu} \, \Bigg[
      \sum_p \, \frac{1}{p^0} \, p^{\mu} \, f^{(0)}_p
      \Bigg], \\
    \label{eq:cond-fit-e-1}
    e &\equiv& u_{\mu} \, \Bigg[
      \sum_p \, \frac{1}{p^0} \, p^{\mu} \, p^{\nu} \, f_p
      \Bigg] \, u_{\nu}
    = u_{\mu} \, \Bigg[
      \sum_p \, \frac{1}{p^0} \, p^{\mu} \, p^{\nu} \, f^{(0)}_p
      \Bigg] \, u_{\nu}.
  \end{eqnarray}
  For consistency,
  one also imposes the constraints to the higher-order terms
  \begin{eqnarray}
    \label{eq:cond-fit-n-2}
    \delta n &=& u_{\mu} \, \Bigg[
      \sum_p \, \frac{1}{p^0} \, p^{\mu} \,\delta f_p 
      \Bigg] = 0, \\
    \label{eq:cond-fit-e-2}
    \delta e &=& u_{\mu} \, \Bigg[
      \sum_p \, \frac{1}{p^0} \, p^{\mu} \, p^{\nu} \, \delta f_p
      \Bigg] \, u_{\nu} = 0.
  \end{eqnarray}
  To obtain the hydrodynamic equation in the particle (Eckart) frame,
  another constraint is imposed
  \begin{eqnarray}
    \nu^\mu = \Delta_{\mathrm{CE}}^{\mu\nu}\, \delta N_\nu =
    \Delta_{\mathrm{CE}}^{\mu\nu} \, \Bigg[
      \sum_p \, \frac{1}{p^0} \, p_{\nu} \, \delta\, f_p
      \Bigg] = 0.
  \end{eqnarray}
  Instead,
  if one wants to obtain the hydrodynamic equation in the energy (Landau-Lifshitz) frame,
  one also imposes the constraint
  \begin{eqnarray}
    \label{eq:cond-fit-landau}
    Q^\mu = \Delta_{\mathrm{CE}}^{\mu\nu}\, \delta T_{\nu\rho} \, u^{\rho} =
    \Delta_{\mathrm{CE}}^{\mu\nu} \, \Bigg[
      \sum_p \, \frac{1}{p^0} \, p_{\nu} \, p_{\rho} \, \delta f_p
      \Bigg]  \, u^{\rho}
    = 0.
  \end{eqnarray}
  The constraints imposed to the distribution function in the higher orders are
  called the \textit{conditions of fit};
  the zeroth-order constraint
  is not a constraint but an identity.
  
  It is noteworthy that although
  a foundation for them has never been given,
  such conditions of fit (\ref{eq:cond-fit-n-1})-(\ref{eq:cond-fit-landau})
  are also imposed in an ad-hoc way
  even when the Maxwell-Grad moment method is adopted \cite{mic001}.
  We stress here that
  these constraints are actually equivalent with
  a strong physical assumption that
  there are no particle-number density nor internal energy of
  the dissipative origin,
  although the distribution function in the non-equilibrium state
  is quite different from that in the local equilibrium state.
  It is, therefore, an urgent but yet unsolved
  problem to verify or elaborate
  these ad-hoc constraints somehow, say, from a microscopic theory,
  or by experiment, if possible.
  
  In the RG method which we adopt,
  one needs no such conditions of fit for derivation,
  and rather the correct forms of them are obtained as a property of the derived equation
  once the frame is specified by the macroscopic-frame vector:
  We shall see that
  the conditions of fit in the energy frame
  is compatible with the underlying Boltzmann equation
  and physical, but those in the particle frame is not
  and thus will never be satisfied in any physical system.
  
  \setcounter{equation}{0}
  \section{
    Reduction of Relativistic Boltzmann Equation with Renormalization-group Method
  }
  \label{sec:2}
  In this section, 
  starting from Eq.(\ref{eq:start})
  we shall derive the relativistic dissipative hydrodynamic equation
  as the infrared asymptotic dynamics
  of the classical relativistic Boltzmann equation
  on the basis of the RG method:
  The five hydrodynamic variables, i.e.,
  the flow velocity, local temperature, and particle-number density (or chemical potential),
  naturally correspond to
  the zero modes of the linearized collision operator.
  Then,
  without recourse to any ansatz
  such as the conditions of fit given
  by Eq.'s (\ref{eq:cond-fit-n-1})-(\ref{eq:cond-fit-landau}),
  the excited modes which are to modify
  the local equilibrium distribution function
  are naturally defined in the sense that they are precisely orthogonal
  to the zero modes with a properly defined inner product for the distribution functions.
  It will be shown on the basis of the inner product
  that the so-called Burnett term does not affect
  the hydrodynamic equation
  owing to the fact that the hydrodynamic modes are
  the zero modes of the linearized collision operator.
  
  \subsection{
    Construction of the approximate solution around arbitrary initial time
  }
  In accordance with the general formulation
  of the RG method \cite{env001,env002,qm,env006}
  \footnote{
    See Appendix B for a brief but self-contained account of the RG method.
  },
  we first try to obtain the perturbative solution $\tilde{f}_p$ 
  to Eq.(\ref{eq:start})
  around the arbitrary initial time $\tau = \tau_0$
  with the initial value $f_p(\tau_0 ,\, \sigma)$;
  \begin{eqnarray}
    \label{eq:1-020}
    \tilde{f}_p(\tau = \tau_0 \,,\, \sigma \,;\, \tau_0) = f_p(\tau_0 \,,\, \sigma),
  \end{eqnarray}
  where we have made explicit that the solution has the $\tau_0$ dependence.
  The initial value is not yet specified,
  we suppose that the initial value is on an exact solution.
  The initial value as well as the solution is expanded
  with respect to $\varepsilon$ as follows;
  \begin{eqnarray}
    \label{eq:1-021}
    \tilde{f}_p(\tau \,,\, \sigma \,;\, \tau_0)
    = \tilde{f}_p^{(0)}(\tau \,,\, \sigma \,;\, \tau_0)
    + \varepsilon \, \tilde{f}_p^{(1)}(\tau \,,\, \sigma \,;\, \tau_0)
    + \varepsilon^2 \, \tilde{f}_p^{(2)}(\tau \,,\, \sigma \,;\, \tau_0)
    + \cdots,
    \nonumber\\
  \end{eqnarray}
  and
  \begin{eqnarray}
    \label{eq:1-022}
    f_p(\tau_0 \,,\, \sigma) = f_p^{(0)}(\tau_0 \,,\, \sigma)
    + \varepsilon \, f_p^{(1)}(\tau_0 \,,\, \sigma)
    + \varepsilon^2 \, f_p^{(2)}(\tau_0 \,,\, \sigma) + \cdots.
  \end{eqnarray}
  The respective initial conditions at $\tau = \tau_0$ are set up as
  \begin{eqnarray}
    \label{eq:1-023}
    \tilde{f}_p^{(l)}(\tau_0 \,,\, \sigma \,;\, \tau_0)
    = f_p^{(l)}(\tau_0 \,,\, \sigma) \,\,\, \mathrm{for} \,\,\, l = 0, \, 1, \, 2,\cdots.
  \end{eqnarray}
  In the expansion, the zeroth-order value
  $\tilde{f}_p^{(0)}(\tau_0 \,,\, \sigma \,;\, \tau_0) = f_p^{(0)}(\tau_0 \,,\, \sigma)$
  is supposed to be as close as possible to an exact solution.
  
  Substituting the above expansions into Eq.(\ref{eq:start})
  in the $\tau$-independent but $\tau_0$-dependent coordinate system with
  \begin{eqnarray}
    \label{eq:1-024}
    \Ren{a}^\mu_p(\tau \,,\, \sigma)
    =
    \Ren{a}^\mu_p(\tau_0 \,,\, \sigma)
    \equiv \Ren{a}^\mu_p(\sigma \,;\, \tau_0),
  \end{eqnarray}
  we obtain the series of the perturbative equations with respect to $\varepsilon$.
  
  Now the zeroth-order equation reads
  \begin{eqnarray}
    \label{eq:1-025}
    \frac{\partial}{\partial \tau} \tilde{f}^{(0)}_p(\tau \,,\, \sigma \,;\, \tau_0)
    = \frac{1}{p \cdot \Ren{a}_p(\sigma \,;\, \tau_0)} \,
    C[\tilde{f}^{(0)}]_p(\tau \,,\, \sigma \,;\, \tau_0).
  \end{eqnarray}
  Since we are interested in the slow motion
  which would be realized asymptotically as $\tau \rightarrow \infty$,
  we should take the following stationary solution or the fixed point,
  \begin{eqnarray}
    \label{eq:1-026}
    \frac{\partial}{\partial \tau}\tilde{f}_p^{(0)}(\tau \,,\, \sigma \,;\, \tau_0) = 0,
  \end{eqnarray}
  which is realized when
  \begin{eqnarray}
    \label{eq:1-027}
    \frac{1}{p \cdot \Ren{a}_p(\sigma \,;\, \tau_0)} \,
    C[\tilde{f}^{(0)}]_p(\tau \,,\, \sigma \,;\, \tau_0) = 0,
  \end{eqnarray}
  for arbitrary $\sigma$.
  Thus, we see that
  $\ln \tilde{f}_p^{(0)}(\tau \,,\, \sigma \,;\, \tau_0)$
  can be represented as a linear combination of the five collision invariants
  $(1 \,,\, p^\mu)$ as mentioned in the last section,
  and hence $\tilde{f}_p^{(0)}(\tau \,,\, \sigma \,;\, \tau_0)$
  is found to be a local equilibrium distribution function
  and thus given by the Juettner function (\ref{eq:Juettner}):
  \begin{eqnarray}
    \label{eq:1-028}
    \tilde{f}_p^{(0)}(\tau \,,\, \sigma \,;\, \tau_0)
    = \frac{1}{(2\pi)^{3}} \,
    \exp \Bigg[ \frac{\mu(\sigma \,;\, \tau_0)
        - p^\mu \, u_\mu(\sigma \,;\, \tau_0)}{T(\sigma \,;\, \tau_0)} \Bigg]
    \equiv f^{\mathrm{eq}}_p(\sigma \,;\, \tau_0).
  \end{eqnarray}
  with $u^\mu(\sigma \,;\, \tau_0) \, u_\mu(\sigma \,;\, \tau_0) = 1$,
  which implies that
  \begin{eqnarray}
    \label{eq:1-029}
    f_p^{(0)}(\tau_0 \,,\, \sigma)
    = \tilde{f}_p^{(0)}(\tau_0 \,,\, \sigma \,;\, \tau_0)
    = f^{\mathrm{eq}}_p(\sigma \,;\, \tau_0).
  \end{eqnarray}
  It should be noticed that the five would-be integration constants
  $T(\sigma \,;\, \tau_0)$, $\mu(\sigma \,;\, \tau_0)$, and $u_\mu(\sigma \,;\, \tau_0)$
  are independent of $\tau$ but may depend on $\tau_0$ as well as $\sigma$.
  In the following,
  we shall suppress the coordinate arguments $(\sigma \,;\, \tau_0)$
  and the momentum subscript, e.g., $p$ when no misunderstanding is expected.
  
  \subsection{
    The linearized collision operator and inner product
  }
  Now the first-order equation reads
  \begin{eqnarray}
    \label{eq:1-030}
    \frac{\partial}{\partial \tau} \tilde{f}_p^{(1)}(\tau)
    = \sum_q \, A_{pq} \, \tilde{f}_q^{(1)}(\tau) + F_p,
  \end{eqnarray}
  where the linear evolution operator $A$ and the inhomogeneous term $F$ are defined by
  \begin{eqnarray}
    \label{eq:1-031}
    A_{pq} &\equiv& \frac{1}{p \cdot \Ren{a}_p} \,
    \frac{\partial}{\partial f_q} C[f]_p \, \Bigg|_{f =
    f^{\mathrm{eq}}}\nonumber\\
    &=& \frac{1}{p \cdot \Ren{a}_p} \,\frac{1}{2!} \, \sum_{p_1} \, \frac{1}{p_1^0} \,
    \sum_{p_2} \, \frac{1}{p_2^0} \, \sum_{p_3} \, \frac{1}{p_3^0} \, 
    \omega(p \,,\, p_1|p_2 \,,\, p_3)\nonumber\\
    & &{} \times ( \delta_{p_2 q} \, f^{\mathrm{eq}}_{p_3}
    + f^{\mathrm{eq}}_{p_2} \, \delta_{p_3 q}
    - \delta_{pq} \, f^{\mathrm{eq}}_{p_1}
    - f^{\mathrm{eq}}_p \, \delta_{p_1 q} ),
  \end{eqnarray}
  and
  \begin{eqnarray}
    \label{eq:1-032}
    F_p \equiv - \frac{1}{p \cdot \Ren{a}_p} \, p \cdot \Ren{\nabla} f^{\mathrm{eq}}_p,
  \end{eqnarray}
  respectively.
  
 To obtain the solution which describes a slow motion,
  it is convenient to first analyze the spectral properties of $A$.
  For this purpose, we convert $A$ to another linear operator
  \begin{eqnarray}
    L \equiv f^{\mathrm{eq}-1} \, A \, f^{\mathrm{eq}},
  \end{eqnarray}
  with the diagonal matrix
  $f^\mathrm{eq}_{pq} \equiv f^\mathrm{eq}_p \, \delta_{pq}$;
  the explicit form of
  $L$ is given by
  \begin{eqnarray}
    \label{eq:1-033}
    L_{pq} =  \frac{-1}{p \cdot \Ren{a}_p} \, \frac{1}{2!} \, \sum_{p_1} \,
    \frac{1}{p_1^0} \, \sum_{p_2} \, \frac{1}{p_2^0} \, \sum_{p_3} \,
    \frac{1}{p_3^0} \, \omega(p \,,\, p_1|p_2 \,,\, p_3) \,
    f^{\mathrm{eq}}_{p_1} \, ( \delta_{p q} + \delta_{p_1 q} -
    \delta_{p_2 q} - \delta_{p_3 q} ).\nonumber\\
  \end{eqnarray}
  Here, we have used the identity
  \begin{eqnarray}
    \label{eq:identity1}
    \omega(p \,,\, p_1|p_2 \,,\, p_3) \, f^{\mathrm{eq}}_{p_2}
    \,f^{\mathrm{eq}}_{p_3} = \omega(p \,,\, p_1|p_2 \,,\, p_3) \, f^{\mathrm{eq}}_{p}
    \,f^{\mathrm{eq}}_{p_1},
  \end{eqnarray}
  which follows from Eq.'s (\ref{eq:1-003}) and (\ref{eq:1-028}).
  
  Let us define the inner product
  between arbitrary non-zero vectors $\varphi$ and $\psi$ by
  \begin{eqnarray}
    \label{eq:1-034}
    \langle  \, \varphi \,,\, \psi \, \rangle
    \equiv \sum_{p} \, \frac{1}{p^0} \, (p \cdot \Ren{a}_p) \,
    f^{\mathrm{eq}}_p \, \varphi_p \, \psi_p.
  \end{eqnarray}
  We note that the norm defined through this inner product is positive definite
  \begin{eqnarray}
    \label{eq:1-035}
    \langle \, \varphi\,,\,\varphi \, \rangle =
    \sum_{p} \, \frac{1}{p^0} \, (p \cdot \Ren{a}_p) \,
    f^{\mathrm{eq}}_p \, (\varphi_p)^2 >
    0\,\,\,\mathrm{for}\,\,\,\varphi_p \ne 0,
  \end{eqnarray}
  since
  \begin{eqnarray}
    \label{eq:1-036}
    p \cdot \Ren{a}_p > 0
  \end{eqnarray}
  in accord with Eq.(\ref{eq:p-a-positive}).
  Notice that the other factors than $ p \cdot \Ren{a}_p$
  in Eq.(\ref{eq:1-035}) are all positive definite.
  We shall see that this positive definiteness (\ref{eq:1-035})
  of the inner product plays an essential role
  in making the resultant hydrodynamic equations
  assure the stability of the thermal equilibrium state,
  as it should be,
  in contrast to some phenomenological equations.
  
  With the inner product,
  it is found that $L$ is self-adjoint
  \begin{eqnarray}
    \label{eq:1-037}
    \langle \, \varphi \,,\, L \, \psi \, \rangle
    &=& - \frac{1}{2!} \, \frac{1}{4} \, 
    \sum_{p} \, \frac{1}{p^0} \,
    \sum_{p_1} \, \frac{1}{p^0_1} \,
    \sum_{p_2} \, \frac{1}{p^0_2} \,
    \sum_{p_3} \, \frac{1}{p^0_3}\nonumber\\
    &&\omega(p \,,\, p_1|p_2 \,,\, p_3) \, f^{\mathrm{eq}}_p \,f^{\mathrm{eq}}_{p_1}\,
    ( \varphi_p + \varphi_{p_1} - \varphi_{p_2} - \varphi_{p_3})
    \, ( \psi_p + \psi_{p_1} - \psi_{p_2} - \psi_{p_3} )\nonumber\\
    &=& \langle \, L \, \varphi \,,\, \psi \, \rangle,
  \end{eqnarray}
  and semi-negative definite
  \begin{eqnarray}
    \label{eq:1-038}
    \langle \, \varphi \,,\, L \, \varphi \, \rangle
    &=&- \frac{1}{2!} \, \frac{1}{4} \, 
    \sum_{p} \, \frac{1}{p^0} \,
    \sum_{p_1} \, \frac{1}{p^0_1} \,
    \sum_{p_2} \, \frac{1}{p^0_2} \,
    \sum_{p_3} \, \frac{1}{p^0_3}\nonumber\\
    &&\times \omega(p \,,\, p_1|p_2 \,,\, p_3) \, f^{\mathrm{eq}}_p \,f^{\mathrm{eq}}_{p_1}\,
    ( \varphi_p + \varphi_{p_1} - \varphi_{p_2} - \varphi_{p_3})^2\nonumber\\
    &\le& 0,
  \end{eqnarray}
  which means that the eigen values of $L$ are zero or negative.
  In the derivation of Eq's (\ref{eq:1-037}) and (\ref{eq:1-038}),
  we have used Eq.'s (\ref{eq:1-003}) and (\ref{eq:identity1}).
  
  The eigen vectors belonging to the zero eigen value are found to be
  \begin{eqnarray}
    \label{eq:1-039}
    \varphi_{0p}^\alpha \equiv \left\{
    \begin{array}{ll}
      \displaystyle{p^\mu} & \displaystyle{\mathrm{for}\,\,\,\alpha = \mu}, \\[2mm]
      \displaystyle{1\times m}     & \displaystyle{\mathrm{for}\,\,\,\alpha = 4},
    \end{array}
    \right.
  \end{eqnarray}
  which span the kernel of $L$ and satisfy
  \begin{eqnarray}
    \label{eq:1-040}
    \big[ L \, \varphi_{0}^\alpha \big]_p = 0.
  \end{eqnarray}
  We call $\varphi_{0}^\alpha$ the \textit{zero modes}.
  It is noted that these zero modes described by the five vectors are
  collision invariants shown in Eq.'s (\ref{eq:1-005}) and
  (\ref{eq:1-006}),  and the factor $m$ in $\varphi_{0p}^4$ is
    introduced merely for convenience
    so that our method can be applied
    to the case of massless particles.

  Following Ref.\citen{env006},
  we define the projection operator $P_0$ onto the kernel of $L$
  which is called the P${}_0$ space and the projection operator $Q_0$
  onto the Q${}_0$ space complement to the P${}_0$ space:
  \begin{eqnarray}
    \label{eq:1-041}
    \big[ P_0 \, \psi \big]_p &\equiv&
    \varphi_{0p}^\alpha \, \eta^{-1}_{0\alpha\beta} \,
    \langle \, \varphi_0^\beta \,,\, \psi \, \rangle,\\
    \label{eq:1-042}
    Q_0 &\equiv& 1 - P_0,
  \end{eqnarray}
  where
  $\eta^{-1}_{0\alpha\beta}$ is the inverse matrix of
  the the P${}_0$-space metric matrix $\eta_0^{\alpha\beta}$
  defined by
  \begin{eqnarray}
    \label{eq:1-043}
    \eta_0^{\alpha\beta} \equiv \langle \, \varphi_0^\alpha \,,\, \varphi_0^\beta \, \rangle.
  \end{eqnarray}
  
  \subsection{
    First-order solution
  }
  The solution to Eq.(\ref{eq:1-030}) with the initial condition
  $\tilde{f}^{(1)}(\tau = \tau_0) = f^{(1)}$, i.e., 
  $\tilde{f}^{(1)}_p(\tau = \tau_0 \,,\, \sigma \,;\, \tau_0) = f^{(1)}_p(\sigma \,;\, \tau_0)$
  is expressed as
  \begin{eqnarray}
    \label{eq:1-044}
    \tilde{f}^{(1)}(\tau) = \mathrm{e}^{(\tau - \tau_0)A} \,
    \Big\{ f^{(1)} + A^{-1} \, \bar{Q}_0 \, F \Big\}
    + (\tau - \tau_0) \, \bar{P}_0 \, F - A^{-1} \, \bar{Q}_0 \, F,
  \end{eqnarray}
  where we have introduced the modified projection operators
  \begin{eqnarray}
    \label{eq:1-045}
    \bar{P}_0 &\equiv& f^{\mathrm{eq}} \, P_0 \, f^{\mathrm{eq}-1},\\
    \label{eq:1-046}
    \bar{Q}_0 &\equiv& f^{\mathrm{eq}} \, Q_0 \, f^{\mathrm{eq}-1}.
  \end{eqnarray}
  We remark that the first term in Eq.(\ref{eq:1-044})
  would be a fast motion coming from the Q${}_0$ space,
  which can be simply eliminated
  by choosing the initial value $f^{(1)}$,
  which has not yet been specified, as
  \begin{eqnarray}
    \label{eq:1-048}
    f^{(1)} = \tilde{f}^{(1)}(\tau_0) = - A^{-1} \, \bar{Q}_0 \, F.
  \end{eqnarray}
  Thus, we have the first-order solution
  \begin{eqnarray}
    \label{eq:1-047}
    \tilde{f}^{(1)}(\tau) = (\tau - \tau_0) \, \bar{P}_0 \, F - A^{-1} \, \bar{Q}_0 \, F,
  \end{eqnarray}
  with the initial value (\ref{eq:1-048}).
  We notice the appearance of the secular term proportional to $\tau - \tau_0$,
  which apparently invalidates the perturbative solution
  when $|\tau - \tau_0|$ becomes large.
  It is worth mentioning that
  the standard Chapman-Enskog expansion method includes a set of conditions
  for not making secular terms appear;
  the conditions are
  the solvability conditions
  of the balance equations (\ref{eq:1-007}) and (\ref{eq:1-008}) \cite{mic001}.
  In applying the solubility conditions,
  one needs apply the ad-hoc constraints
  on the distribution function
  for defining the flow (eg. Eckart flow or Landau-Lifshitz flow)
  as well as the ad-hoc constraints on the particle-number density
  and internal energy \cite{mic001},
  as given by Eq.'s(\ref{eq:cond-fit-n-1})-(\ref{eq:cond-fit-landau}).
  In the present RG method,
  secular terms are allowed to appear
  and no constraints are imposed on the distribution function;
  rather the secular terms will be utilized to obtain the slow dynamics.
  
  We remark here that we could apply the RG method here to Eq.(\ref{eq:1-047}),
  which will give the relativistic Euler equation
  without dissipation effects.
  To get a dissipative hydrodynamic equation,
  we need to proceed to the second order in our method.
  
  \subsection{
    Second-order solution
  }
  The second-order equation is written as
  \begin{eqnarray}
    \label{eq:1-049}
    \frac{\partial}{\partial \tau} \tilde{f}^{(2)}_p(\tau)
    = \sum_q A_{pq} \, \tilde{f}^{(2)}_q(\tau)
    + (\tau - \tau_0)^2 \, G_p + (\tau - \tau_0) \, H_p + I_p,
  \end{eqnarray}
  where
  \begin{eqnarray}
    \label{eq:1-050}
    B_{pqr} &\equiv& \frac{1}{p \cdot \Ren{a}_p} \,
    \frac{1}{2} \, \frac{\partial^2}{\partial f_q \partial f_r} C[f]_p
    \, \Bigg|_{f = f^{\mathrm{eq}}}\nonumber\\
    &=& \frac{1}{p \cdot \Ren{a}_p} \,\frac{1}{2}\,\frac{1}{2!} \, \sum_{p_1} \, \frac{1}{p_1^0} \,
    \sum_{p_2} \, \frac{1}{p_2^0} \, \sum_{p_3} \, \frac{1}{p_3^0} \, 
    \omega(p \,,\, p_1|p_2 \,,\, p_3)\nonumber\\
    && \times ( \delta_{p_2 q} \, \delta_{p_3 r}
    + \delta_{p_2 r} \, \delta_{p_3 q}
    - \delta_{pq} \, \delta_{p_1 r}
    - \delta_{p r} \, \delta_{p_1 q} ),\\
    \label{eq:1-051}
    G_p &\equiv& \sum_q \, \sum_r \, B_{pqr} \,
    \big[ \bar{P}_0 \, F \big]_q \, \big[ \bar{P}_0 \, F \big]_r,\\
    \label{eq:1-052}
    H_p &\equiv& - \sum_q \, \sum_r \, B_{pqr} \,
    \Big( \big[ \bar{P}_0 \, F \big]_q \, \big[ A^{-1} \, \bar{Q}_0 \, F \big]_r
    + \big[ A^{-1} \, \bar{Q}_0 \, F \big]_q \, \big[ \bar{P}_0 \, F \big]_r \Big)\nonumber\\
    &&{}- \frac{1}{p \cdot \Ren{a}_p} \, p \cdot \Ren{\nabla} \big[ \bar{P}_0 \, F \big]_p,\\
    \label{eq:1-053}
    I_p &\equiv& \sum_q \, \sum_r \, B_{pqr} \,
    \big[ A^{-1} \, \bar{Q}_0 \, F \big]_q \, \big[ A^{-1} \, \bar{Q}_0 \, F \big]_r
    + \frac{1}{p \cdot \Ren{a}_p} \, p \cdot \Ren{\nabla}
    \big[ A^{-1} \, \bar{Q}_0 \, F \big]_p.
  \end{eqnarray}
  The solution to Eq.(\ref{eq:1-049}) is found to be
  \begin{eqnarray}
    \label{eq:1-054}
    \tilde{f}^{(2)}(\tau) &=& \mathrm{e}^{(\tau - \tau_0)A} \,
    \Big\{ f^{(2)} + 2 \, A^{-3} \, \bar{Q}_0 \, G
    + A^{-2} \, \bar{Q}_0 \, H  + A^{-1} \, \bar{Q}_0 \, I \Big\}\nonumber\\
    & &{} + \frac{1}{3} \, (\tau - \tau_0)^3 \, \bar{P}_0 \, G
    + \frac{1}{2} \, (\tau - \tau_0)^2 \, \Big\{ \bar{P}_0 \, H
    - 2 \, A^{-1} \, \bar{Q}_0 \, G \Big\}\nonumber\\
    & &{} + (\tau - \tau_0) \, \Big\{ \bar{P}_0 \, I
    - 2 \, A^{-2} \, \bar{Q}_0 \, G - A^{-1} \, \bar{Q}_0 \, H \Big\}\nonumber\\
    &&{}+ \Big\{ - 2 \, A^{-3} \, \bar{Q}_0 \, G
    - A^{-2} \, \bar{Q}_0 \, H - A^{-1} \, \bar{Q}_0 \, I \Big\}.
  \end{eqnarray}
  Again the would-be fast motion can be eliminated
  by a choice of the initial value $f^{(2)}$ so that
  \begin{eqnarray}
    \label{eq:1-056}
    f^{(2)} = \tilde{f}^{(2)}(\tau_0) =
    - 2 \, A^{-3} \, \bar{Q}_0 \, G - A^{-2} \, \bar{Q}_0 \, H - A^{-1} \, \bar{Q}_0 \, I.
  \end{eqnarray}
  Then, we have a second-order solution
  \begin{eqnarray}
    \label{eq:1-055}
    \tilde{f}^{(2)}(\tau) &=& \frac{1}{3} \, (\tau - \tau_0)^3 \, \bar{P}_0 \, G
    + \frac{1}{2} \, (\tau - \tau_0)^2 \, \Big\{ \bar{P}_0 \, H
    - 2 \, A^{-1} \, \bar{Q}_0 \, G \Big\}\nonumber\\
    & &{} + (\tau - \tau_0) \, \Big\{ \bar{P}_0 \, I
    - 2 \, A^{-2} \, \bar{Q}_0 \, G - A^{-1} \, \bar{Q}_0 \, H \Big\}\nonumber\\
    &&{}+ \Big\{ - 2 \, A^{-3} \, \bar{Q}_0 \, G - A^{-2} \, \bar{Q}_0 \, H
    - A^{-1} \, \bar{Q}_0 \, I \Big\},
  \end{eqnarray}
  We notice again
  the appearance of secular terms
  and that no constraints on the solution
  are imposed for defining the flow,
  in contrast to the standard Chapman-Enskog expansion method \cite{mic001}.
  
  Summing up the perturbative solutions up to the second order,
  we have an approximate solution around $\tau \simeq \tau_0$ to this order;
  \begin{eqnarray}
    \label{eq:1-057}
    \tilde{f}_p(\tau \,,\, \sigma \,;\, \tau_0)
    = \tilde{f}^{(0)}_p(\tau \,,\, \sigma \,;\, \tau_0)
    + \varepsilon \, \tilde{f}^{(1)}_p(\tau \,,\, \sigma \,;\, \tau_0)
    + \varepsilon^2 \, \tilde{f}^{(2)}_p(\tau \,,\, \sigma \,;\, \tau_0)
    + O(\varepsilon^3).\nonumber\\
  \end{eqnarray}
  We emphasize that this solution contains
  the secular terms which apparently invalidates the perturbative expansion
  for $\tau$ away from the initial time $\tau_0$.
  
  \subsection{
    RG improvement of perturbative expansion
  }
  The point of the RG method lies in the fact that
  we can utilize the secular terms to obtain
  an asymptotic solution valid in a global domain.
  Now we may see that
  we have a family of curves
  $\tilde{f}_p(\tau \,,\, \sigma \,;\, \tau_0)$
  parameterized with $\tau_0$.
  They are all on the exact solution
  $f_p(\sigma \,;\, \tau)$ at $\tau = \tau_0$ up to $O(\varepsilon^3)$,
  but only valid locally for $\tau$ near $\tau_0$.
  So it is conceivable that the envelope of the family of curves
  which contacts with each local solution at $\tau = \tau_0$ will
  give a global solution in our asymptotic situation.
  According to the classical theory of envelopes,
  the envelope which contact with any curve in the family
  at $\tau = \tau_0$ is obtained by
  \footnote{
    See Appendix B for the foundation of this procedure.
  }
  \begin{eqnarray}
    \label{eq:1-058}
    \frac{\mathrm{d}}{\mathrm{d}\tau_0}
    \tilde{f}_p(\tau \,,\, \sigma \,;\, \tau_0) \Bigg|_{\tau_0 = \tau} = 0,
  \end{eqnarray}
  or explicitly
  \begin{eqnarray}
    \label{eq:1-059}
    &&\frac{\partial}{\partial \tau} \Big\{ f^{\mathrm{eq}}
    - \varepsilon \, A^{-1} \, \bar{Q}_0 \, F \Big\}
    - \varepsilon \, \bar{P}_0 \, F\nonumber\\
    &&\hspace{2cm}{}- \varepsilon^2 \, \Big\{ \bar{P}_0 \, I
    - 2 \, A^{-2} \, \bar{Q}_0 \, G
    - A^{-1} \, \bar{Q}_0 \, H \Big\} + O(\varepsilon^3) = 0.\nonumber\\
  \end{eqnarray}
  This envelope equation is the basic equation in the RG method
  and gives the equation of motion governing the dynamics of
  the five slow variables
  $T(\sigma \,;\, \tau)$, $\mu(\sigma \,;\, \tau)$
  and $u^\mu(\sigma \,;\, \tau)$
  in $f^\mathrm{eq}_p(\sigma \,;\, \tau)$.
  The global solution in the asymptotic region
  is given as an envelope function,
  \begin{eqnarray}
    \label{eq:1-060}
    f_{\mathrm{E}p}(\tau\,,\,\sigma) &\equiv &
    \tilde{f}_p(\tau\,,\,\sigma \,;\, \tau_0 = \tau)\nonumber \\
    &=& f^{\mathrm{eq}}_p(\sigma \,;\, \tau)
    - \varepsilon \, \big[ A^{-1} \, \bar{Q}_0 \, F \big]_p(\sigma \,;\, \tau)\nonumber\\
    & &{} - \varepsilon^2 \, \Big\{ 2 \, \big[ A^{-3} \, \bar{Q}_0 \, G
    \big]_p(\sigma \,;\, \tau)\nonumber\\
    &&{}
    + \big[ A^{-2} \, \bar{Q}_0 \, H \big]_p(\sigma \,;\, \tau)
    + \big[ A^{-1} \, \bar{Q}_0 \, I \big]_p(\sigma \,;\, \tau) \Big\} + O(\varepsilon^3),
  \end{eqnarray}
  where the exact solution of Eq.(\ref{eq:1-059}) is inserted.
  As is proved in Appendix B,
  the envelope function $f_{\mathrm{E}p}(\tau\,,\,\sigma)$ satisfies Eq.(\ref{eq:start})
  in a global domain up to $O(\varepsilon^3)$ owing to the condition (\ref{eq:1-058}),
  although $\tilde{f}_p(\tau \,,\, \sigma \,;\, \tau_0)$ itself was constructed
  as a local solution around $\tau\sim \tau_0$.
  Thus, one sees that $f_{\mathrm{E}p}(\tau\,,\,\sigma)$ now describes
  a coarse-grained evolution
  of the one-particle distribution function in Eq.(\ref{eq:start}),
  because the time-derivatives of the quantities in $f_{\mathrm{E}p}(\tau\,,\,\sigma)$ are all
  in the order of $\varepsilon$ or higher.
  We emphasize that
  we have derived the slow-motion equation of Eq.(\ref{eq:start})
  in the form of the pair of Eq.'s (\ref{eq:1-059}) and (\ref{eq:1-060}).

  \subsection{
    Reduction of the RG equation to generic hydrodynamic equation
  }
  Now let us see that
  the RG/Envelope equation (\ref{eq:1-059}) is actually
  the hydrodynamic equation
  governing the five slow variables,
  $T(\sigma \,;\, \tau)$, $\mu(\sigma \,;\, \tau)$, and $u^\mu(\sigma \,;\, \tau)$.
  To show this explicitly,
  we apply $\bar{P}_0$ from the left and then take the inner product with
  the five zero modes $\varphi_0^\alpha$.
  In this procedure,
  we first note that
  the direct use of the definitions of
  Eq.'s (\ref{eq:1-041}), (\ref{eq:1-043}), and (\ref{eq:1-045})
  leads to the following identity;
  \begin{eqnarray}
    \label{eq:1-062}
    \sum_{p} \, \frac{1}{p^0} \, (p \cdot \Ren{a}_p) \,
    \varphi_{0p}^{\alpha} \, \big[ \bar{P}_0 \, \psi \big]_p
    = \sum_{p} \, \frac{1}{p^0} \, (p \cdot \Ren{a}_p) \, \varphi_{0p}^{\alpha} \, \psi_p.
  \end{eqnarray}
  Furthermore,
  noting that
  $\varphi^\alpha_{0p}$ are the collision invariants
  as shown in Eq.'s (\ref{eq:1-005}) and (\ref{eq:1-006}),
  we have
  \begin{eqnarray}
    \label{eq:1-063}
    \sum_{p} \, \frac{1}{p^0} \, (p \cdot \Ren{a}_p) \,
    \varphi_{0p}^\alpha \, \sum_{q} \, \sum_{r} \, B_{pqr} \,
    \big[ A^{-1} \, \bar{Q}_0 \, F \big]_q \, \big[ A^{-1} \, \bar{Q}_0 \, F \big]_r
    = 0.
  \end{eqnarray}
  Here, we have used the relation
  \begin{eqnarray}
    \sum_{q} \, \sum_{r} \, B_{pqr} \,
    \big[ A^{-1} \, \bar{Q}_0 \, F \big]_q \, \big[ A^{-1} \, \bar{Q}_0 \, F \big]_r
    = \frac{1}{p\cdot\Ren{a}_p} \, C[ A^{-1} \, \bar{Q}_0 \, F ]_p,
  \end{eqnarray}
  which follows from
  the definitions of Eq.'s (\ref{eq:1-002}) and (\ref{eq:1-050}).
  
  Thus, we have
  \begin{eqnarray}
    \label{eq:1-061}
    \sum_{p} \, \frac{1}{p^0} \, \varphi_{0p}^\alpha \,
    \Bigg[ (p \cdot \Ren{a}_p) \, \frac{\partial}{\partial \tau}
      + \varepsilon \, p \cdot \Ren{\nabla} \Bigg]
    \Big\{ f^{\mathrm{eq}}_p
    - \varepsilon \, \big[ A^{-1} \, \bar{Q}_0 \, F \big]_p \Big\} + O(\varepsilon^3) = 0,
  \end{eqnarray}
  Putting back $\varepsilon = 1$,
  we arrive at
  \begin{eqnarray}
    \label{eq:1-064}
    \partial_\mu J^{\mu\alpha}_{\mathrm{1st}} = 0,
  \end{eqnarray}
  with
  \begin{eqnarray}
    \label{eq:1-065}
    J^{\mu\alpha}_{\mathrm{1st}}
    \equiv \sum_{p} \, \frac{1}{p^0} \, p^\mu \, \varphi_{0p}^\alpha \,
    \Big\{ f^{\mathrm{eq}}_p - \big[ A^{-1} \, \bar{Q}_0 \, F \big]_p \Big\},
  \end{eqnarray}
  where we have used
  \begin{eqnarray}
    \label{eq:1-066}
    (p \cdot \Ren{a}_p) \, \frac{\partial}{\partial \tau}
    + p \cdot \Ren{\nabla} = p^\mu\,\partial_\mu.
  \end{eqnarray}
  It is noted that
  $J^{\mu\alpha}_{\mathrm{1st}}$ perfectly agrees with
  the one obtained by inserting the solution
  $f_{\mathrm{E}p}(\tau\,,\,\sigma)$
  in Eq.(\ref{eq:1-060})
  into $N^\mu$ and $T^{\mu\nu}$ in Eq.'s (\ref{eq:1-007}) and (\ref{eq:1-008}):
  \begin{eqnarray}
    \label{eq:1-067}
    N^{\mu} &=& m^{-1} \, J^{\mu 4}_{\mathrm{1st}},\\
    \label{eq:1-068}
    T^{\mu\nu} &=& J^{\mu\nu}_{\mathrm{1st}}.
  \end{eqnarray}
  Therefore,
  we conclude that Eq.(\ref{eq:1-064})
  is identically the relativistic dissipative hydrodynamic equation:
  We note that
  the subscript of ``1st'' in $J^{\mu\alpha}_{\mathrm{1st}}$
  means that the obtained equation is
  in a class of
  the so-called \textit{first-order} relativistic dissipative hydrodynamics.
  As is shown in Appendix A,
  the first-order equations can be derived phenomenologically
  with use of the entropy current
  which includes dissipative effects up to the first order.
  
  It is noted that
  the hydrodynamic equation (\ref{eq:1-064}) with the currents (\ref{eq:1-065})
  still contains the macroscopic-frame vector 
  $\Ren{a}_p^\mu$,
  which is now dependent on $\tau$ as well as on $\sigma$.
  We shall see that Eq.(\ref{eq:1-065}) reduces to the currents in various frames with 
  a choice of $\Ren{a}_p^\mu$.
  In other words,
  we have obtained the relativistic hydrodynamic equations for a
  viscous fluid in a generic frame,
  which is a kind of the master equation
  from which various relativistic hydrodynamic equations are deduced.
  As far as we are aware of,
  this is the first time when such a generic hydrodynamic
  equation for the relativistic viscous system is obtained.
  This is one of the main results in the present work.
  We stress that
  this was made possible
  because any ad-hoc ansatz such as the conditions of fit is not
  necessary in the RG method.
  
  It is also worth mentioning that
  the problematic Burnett term is absent in Eq.(\ref{eq:1-061})
  thanks to Eq.(\ref{eq:1-063}).
  If the Burnett term were to remain,
  the particle-number and energy-momentum conservation laws are lost;
  moreover,
  boundary conditions might have to be taken care of simultaneously
  because its magnitude is comparable to the Burnett term \cite{landau}.
  In fact,
  the presence of the Burnett term is known to be inevitable
  when the Chapman-Enskog expansion method \cite{landau}
  is applied to derive the Navier-Stokes equation
  from the non-relativistic Boltzmann equation
  \footnote{
    We did not recognize this important point in Ref.\citen{env009}
    that the RG method naturally gives
    a hydrodynamic equation free from the Burnett term.
    We expect that the mechanism similar to Eq.(\ref{eq:1-063})
    is realized in the non-relativistic case.
  }.
  
  We decompose $J^{\mu\alpha}_{\mathrm{1st}}$ into two parts as
  $J^{\mu\alpha}_{\mathrm{1st}} = J^{(0)\mu\alpha}_{\mathrm{1st}}
  + \delta J^{\mu\alpha}_{\mathrm{1st}}$,
  where
  \begin{eqnarray}
    \label{eq:1-069}
    J^{(0)\mu\alpha}_{\mathrm{1st}} &\equiv&
    \sum_{p} \, \frac{1}{p^0} \, p^\mu \, \varphi_{0p}^\alpha \, f^{\mathrm{eq}}_p,\\
    \label{eq:1-070}
    \delta J^{\mu\alpha}_{\mathrm{1st}} &\equiv&
    - \sum_{p} \, \frac{1}{p^0} \, p^\mu \, \varphi_{0p}^\alpha \,
    \big[A^{-1} \, \bar{Q}_0 \, F \big]_p
    = - \langle\,\tilde{\varphi}^{\mu\alpha}_1\,,\,L^{-1}\,Q_0\,f^{\mathrm{eq}-1}\,F \,\rangle,
  \end{eqnarray}
  with
  \begin{eqnarray}
    \label{eq:1-071}
    \tilde{\varphi}^{\mu\alpha}_{1p} \equiv
    p^\mu \, \varphi_{0p}^\alpha \, \frac{1}{p \cdot \Ren{a}_p}.
  \end{eqnarray}
  Needless to say,
  $J^{(0)\mu\alpha}_{\mathrm{1st}}$ and
  $\delta J^{\mu\alpha}_{\mathrm{1st}}$ represent
  the currents in the perfect-fluid and
  dissipative part, respectively.
  Corresponding to 
  $J^{(0)\mu\alpha}_{\mathrm{1st}}$ and
  $\delta J^{\mu\alpha}_{\mathrm{1st}}$,
  $N^\mu$ and $T^{\mu\nu}$ in Eq.'s (\ref{eq:1-067}) and (\ref{eq:1-068})
  are decomposed as
  \begin{eqnarray}
    N^{\mu} &=& N^{(0)\mu} + \delta N^\mu, \\
    T^{\mu\nu} &=& T^{(0)\mu\nu} + \delta T^{\mu\nu},
  \end{eqnarray}
  where
  \begin{eqnarray}
    \label{eq:N-perfect}
    N^{(0)\mu} &\equiv& m^{-1} \, J^{(0)\mu 4}_{\mathrm{1st}},\\
    \label{eq:N-dissipative}
    \delta N^{\mu} &\equiv& m^{-1} \, \delta J^{\mu 4}_{\mathrm{1st}},\\
    \label{eq:T-perfect}
    T^{(0)\mu\nu} &\equiv& J^{(0)\mu\nu}_{\mathrm{1st}},\\
    \label{eq:T-dissipative}
    \delta T^{\mu\nu} &\equiv& \delta J^{\mu\nu}_{\mathrm{1st}}.
  \end{eqnarray}
  
  For later convenience,
  we present a simpler form of $\delta J^{\mu\alpha}_{\mathrm{1st}}$.
  First, we note that $F$ in Eq.(\ref{eq:1-032}) is expressed as
  \begin{eqnarray}
    \label{eq:1-072}
    F_p = - f^\mathrm{eq}_p \, \Big(\tilde{\varphi}^{\mu4}_{1p} \,
    m^{-1} \,
    \Ren{\nabla}_\mu \frac{\mu}{T} - \tilde{\varphi}^{\mu\nu}_{1p} \,
    \Ren{\nabla}_\mu \frac{u_\nu}{T}\Big)
    = - f^\mathrm{eq}_p \, \tilde{\varphi}_{1p}^{\mu\alpha} \, \bar{X}_{\mu\alpha},
  \end{eqnarray}
  where
  \begin{eqnarray}
    \label{eq:1-073}
    \bar{X}_{\mu\alpha} \equiv \left\{
    \begin{array}{ll}
      - \Ren{\nabla}_\mu (u_\nu / T) & \displaystyle{\mathrm{for}\,\,\,\alpha = \nu}, \\[2mm]
      m^{-1} \, \Ren{\nabla}_\mu (\mu /T) & \displaystyle{\mathrm{for}\,\,\,\alpha = 4}.
    \end{array}
    \right.
  \end{eqnarray}
  Using the above representation of $F$ and the identity
  \begin{eqnarray}
    \label{eq:1-074}
    \langle \, \varphi \,,\, L^{-1}\,Q_0 \, \psi \, \rangle = \langle \, Q_0 \,
    \varphi \,,\, L^{-1}\,Q_0 \, \psi \, \rangle,
  \end{eqnarray}
  we can reduce the dissipative part $\delta J^{\mu\alpha}_{\mathrm{1st}}$
  to the following form:
  \begin{eqnarray}
    \label{eq:1-075}
    \delta J^{\mu\alpha}_\mathrm{1st}
    = \eta^{\mu\alpha\nu\beta}_1 \, \bar{X}_{\nu\beta},
  \end{eqnarray}
  where
  \begin{eqnarray}
    \label{eq:1-076}
    \eta^{\mu\alpha\nu\beta}_1 \equiv \langle
    \, \varphi^{\mu\alpha}_1\,,\,
    L^{-1} \, \varphi^{\nu\beta}_1 \, \rangle.
  \end{eqnarray}
  Here, we have introduced an important new vector defined by
  \begin{eqnarray}
    \label{eq:1-077}
    \varphi^{\mu\alpha}_{1p} \equiv \big[ Q_0 \,
      \tilde{\varphi}^{\mu\alpha}_1 \big]_p.
  \end{eqnarray}
  We call $\varphi^{\mu\alpha}_{1p}$ the \textit{first-excited modes}.
  It is noteworthy that
  $\delta J^{\mu\alpha}_{\mathrm{1st}}$
  is represented as a product of
  $\eta^{\mu\alpha\nu\beta}_1$ and $\bar{X}_{\nu\beta}$:
  $\eta^{\mu\alpha\nu\beta}_1$ has some information
  about the transport coefficients,
  while $\bar{X}_{\nu\beta}$ is identical to the corresponding thermodynamic forces.
  
  Finally,
  we write down the relativistic dissipative hydrodynamic equation
  defined in the covariant coordinate system $(\sigma\,;\,\tau)$:
  \begin{eqnarray}
    \label{eq:1-078}
    \sum_{p} \, \frac{1}{p^0} \, \varphi_{0p}^\alpha \,
    \Bigg[ (p \cdot \Ren{a}_p) \, \frac{\partial}{\partial \tau}
      + p \cdot \Ren{\nabla} \Bigg]
    \Bigg[ f^{\mathrm{eq}}_p \, \Big(1 +
    \big[ L^{-1} \, \varphi^{\nu\beta}_1 \big]_p \,
    \bar{X}_{\nu\beta} \Big) \Bigg] = 0,
  \end{eqnarray}
  which can be derived straightforwardly
  from Eq.'s (\ref{eq:1-061}) and (\ref{eq:1-072}).
  This form of the relativistic dissipative hydrodynamic equation
  will be found to play an essential role
  in the stability analysis to be presented in \S\ref{sec:6}.
  
  \setcounter{equation}{0}
  \section{
    Relativistic Hydrodynamic Equation
    with Specific Macroscopic-frame Vectors
  }
  \label{sec:3}
  In this section,
  we give explicit forms of 
  the relativistic dissipative hydrodynamic equations
  by integrating out
  the right-hand side of Eq.'s (\ref{eq:1-069}) and (\ref{eq:1-070})
  with respect to the momentum $p^\mu$.
  
  A remark is in order here:
  We calculate and present the thermodynamic quantities and transport coefficients
  using our model equation, i.e., the relativistic Boltzmann equation, for completeness.
  Then,
  the explicit forms of them are inherently for the relativistic rarefied gas.
  We would like to remind the reader, however, that
  the main purpose of the present work is to determine
  the form of the relativistic hydrodynamic equations for a viscous fluid,
  and
  expect that the forms of the macroscopic hydrodynamic equations
  which contain the thermodynamic quantities and transport coefficients
  only parametrically,
  and hence the forms are
  independent of the microscopic expressions of these quantities.
  
  After the integration, the currents of perfect-fluid part
  $J^{(0)\mu\alpha}_\mathrm{1st}$ reads
  \begin{eqnarray}
    \label{eq:2-001}
    J^{(0)\mu\alpha}_\mathrm{1st} = \left\{
    \begin{array}{ll}
      \displaystyle{
        e \, u^\mu \, u^\nu - p \, \Delta^{\mu\nu} = T^{(0)\mu\nu}
      }
      & \displaystyle{\mathrm{for}\,\,\,\alpha = \nu,} \\[2mm]
      \displaystyle{
        m \, n \, u^\mu = m\,N^{(0)\mu}
      }
      & \displaystyle{\mathrm{for}\,\,\,\alpha = 4,}
    \end{array}
    \right.
  \end{eqnarray}
  where
  $e$, $p$, and $n$
  denote
  the internal energy,
  the pressure,
  and the particle-number density,
  respectively.
  These quantities are defined by
  \begin{eqnarray}
    \label{eq:2-002}
    n &\equiv& \sum_{p} \, \frac{1}{p^0} \,
    f^{\mathrm{eq}}_p \, (p \cdot u)
    = (2\pi)^{-3} \, 4\pi \, m^3 \, \mathrm{e}^{\frac{\mu}{T}}
    \, z^{-1} \, K_2(z),\\
    \label{eq:2-003}
    e &\equiv& \sum_{p} \, \frac{1}{p^0} \, f^{\mathrm{eq}}_p \, (p
    \cdot u)^2 = m \, n \, \Bigg[
      \frac{K_3(z)}{K_2(z)} - z^{-1} \Bigg],\\
    \label{eq:2-004}
    p &\equiv& \sum_{p} \, \frac{1}{p^0} \, f^{\mathrm{eq}}_p \, (-1/3\,p^\mu
    \, p^\nu \, \Delta_{\mu\nu}) = n \, T,
  \end{eqnarray}
  where
  we have introduced
  the second- and third-order modified Bessel functions
  $K_2(z)$ and $K_3(z)$ with $z$ being
  the dimensionless variable defined by
  \begin{eqnarray}
    \label{eq:2-005}
    z \equiv \frac{m}{T}.
  \end{eqnarray}
  The explicit form of the modified Bessel functions
  is presented in Eq.(\ref{eq:6-008}) in Appendix C.
  
  \subsection{
    A generic choice of the macroscopic-frame vector
  }
  The dissipative part of the currents
  $\delta J^{\mu\alpha}_\mathrm{1st}$
  depends on the macroscopic-frame vector $\Ren{a}^\mu_p$ explicitly,
  in the contrast of $J^{(0)\mu\alpha}_\mathrm{1st}$.
  Here, we shall consider $\delta J^{\mu\alpha}_\mathrm{1st}$
  with a choice of $\Ren{a}_p^\mu$.
  As a simple but nontrivial choice,
  let us take the following set
  of the macroscopic-frame vectors with $\theta$ being a constant;
  \begin{eqnarray}
    \label{eq:2-006}
    \Ren{a}_p^\mu = \frac{1}{p \cdot u} \, \Big( (p \cdot u) \, \cos\theta
    + m \, \sin\theta \Big) \, u^\mu \equiv \theta_p^\mu.
  \end{eqnarray}
  In Eq.(\ref{eq:2-006}),
  we note that
  the factor $m$ is introduced
  simply to make the expression dimensionless,
  so our method is also applicable
  to the case of massless particles.
  
  A remark is in order here:
  The inequality (\ref{eq:1-013-2}) of $\Ren{a}^0_p$
  leads to a restriction on $\theta$
  when $\Ren{a}^\mu_p = \theta^\mu_p$ in Eq.(\ref{eq:1-036}),
  as
  $(p \cdot u) \, \cos\theta  + m \, \sin\theta\, 
  =\sqrt{ (p \cdot u)^2 + m^2}\, \sin (\theta +\chi) > 0$
  with $\tan \chi = (p\cdot u) / m \ge 1$,
  which implies that
  \begin{eqnarray}
    \label{eq:2-007}
    -\frac{\pi}{4}<  \theta \le \frac{\pi}{2},
  \end{eqnarray}
  because $\pi/4 \le \chi < \pi/2$.
  
  Under these settings,
  we shall now give
  the explicit representation of
  $\delta J^{\mu\alpha}_\mathrm{1st} = \eta^{\mu\alpha\nu\beta}_1 \, \bar{X}_{\nu\beta}$.
  First,
  we consider $\bar{X}_{\mu\alpha}$ in Eq.(\ref{eq:1-075}).
  Inserting the equality $\Ren{a}_p^\mu = \theta_p^\mu$, we have
  \begin{eqnarray}
    \label{eq:2-008}
    \Ren{\Delta}_p^{\mu\nu} = g^{\mu\nu} - u^\mu \, u^\nu = \Delta^{\mu\nu},
  \end{eqnarray}
  so that the differential operator with respect to $\sigma^\mu$ reads
  \begin{eqnarray}
    \label{eq:2-009}
    \Ren{\nabla}^\mu = \Delta^{\mu\nu} \, \partial_\nu \equiv \nabla^\mu.
  \end{eqnarray}
  Notice that
  $\Delta^{\mu\nu}$ and $\nabla^\mu$ are now identical
  to the familiar projection matrix onto the subspace complement to $u^\mu$
  and the covariant spatial differential operator \cite{mic001}, respectively.
  Thus, the thermodynamic forces $\bar{X}_{\mu\alpha}$ are reduced to
  \begin{eqnarray}
    \label{eq:2-010}
    \bar{X}_{\mu\alpha} = \left\{
    \begin{array}{ll}
      - \nabla_\mu (u_\nu / T) & \displaystyle{\mathrm{for}\,\,\,\alpha = \nu}, \\[2mm]
      m^{-1} \, \nabla_\mu (\mu /T) & \displaystyle{\mathrm{for}\,\,\,\alpha = 4}.
    \end{array}
    \right.
  \end{eqnarray}
  
  Next, we derive an explicit form of $\eta^{\mu\alpha\nu\beta}_1$ in Eq.(\ref{eq:1-075}),
  which task is tantamount to obtaining an explicit representation
  of $\varphi^{\mu\alpha}_{1p}$.
  After a straightforward calculation presented in Appendix \ref{sec:8},
  one can find that
  $\varphi^{\mu\alpha}_{1p}$ are written as
  \begin{eqnarray}
    \label{eq:2-011}
    \varphi^{\mu\alpha}_{1p}
    = \left\{
    \begin{array}{ll}
      \displaystyle{
        \tilde{\Pi}_p \, \Big(Y_1(\theta) \, u^\mu \, u^\nu - Y_2(\theta) \, \Delta^{\mu\nu}\Big)
      }
      & \displaystyle{} \\[2mm]
      \displaystyle{
        \hspace{2cm}
        {}+ \tilde{J}^\mu_p \, Y_3(\theta) \, u^\nu + \tilde{J}^\nu_p \, Y_3(\theta) \, u^\mu
        + \tilde{\pi}^{\mu\nu}_p
      }
      & \displaystyle{\mathrm{for}\,\,\,\alpha = \nu,} \\[2mm]
      \displaystyle{
        \tilde{\Pi}_p \, Z_1(\theta) \, u^\mu + \tilde{J}^\mu_p \, Z_2(\theta)
      }
      & \displaystyle{\mathrm{for}\,\,\,\alpha = 4.}
    \end{array}
    \right.\nonumber\\
  \end{eqnarray}
  Here, we have introduced
  $\tilde{\Pi}_p$, $\tilde{J}^\mu_p$, and $\tilde{\pi}^{\mu\nu_p}$
  defined by
  \begin{eqnarray}
    \label{eq:2-012}
    (\tilde{\Pi}_p,\,\tilde{J}^\mu_p,\,\tilde{\pi}^{\mu\nu}_p)
    &\equiv& (\Pi_p,\,J^\mu_p,\,\pi^{\mu\nu}_p)
    \, \frac{1}{p\cdot \theta_p},\\
    \label{eq:2-013}
    \Pi_p &\equiv& \Big( \frac{4}{3} - \gamma \Big) \, (p \cdot u)^2
    + \Big( (\gamma - 1) \, T \, \hat{h} - \gamma \, T \Big) \, (p \cdot u)
    - \frac{1}{3} \, m^2,\\
    \label{eq:2-014}
    J^\mu_p &\equiv& - \Big( (p \cdot u) - T \, \hat{h} \Big) \, \Delta^{\mu\nu} \, p_\nu,\\
    \label{eq:2-015}
    \pi^{\mu\nu}_p &\equiv& \Delta^{\mu\nu\rho\sigma} \, p_\rho \,
    p_\sigma,
  \end{eqnarray}
  and $Y_1(\theta)$, $Y_2(\theta)$, $Y_3(\theta)$, $Z_1(\theta)$, and $Z_2(\theta)$
  defined by
  \begin{eqnarray}
    \label{eq:2-016}
    Y_1(\theta) &\equiv&
    \frac{
      - 3 \, z^2 \, \sin^2\theta
    }{
      z^2 \, \cos^2\theta + z^2 \, (3 \, \gamma - 4) \,
      \sin^2\theta
      - 3 \, z \, \big[ 1 - (\hat{h} - 1) \, (\gamma - 1) \big] \,
      \cos\theta \, \sin\theta
    }
    ,\nonumber\\\\
    \label{eq:2-017}
    Y_2(\theta) &\equiv&
    \frac{
      z^2 \, \cos^2\theta - z^2 \, \sin^2\theta
    }{
      z^2 \, \cos^2\theta + z^2 \, (3 \, \gamma - 4) \,
      \sin^2\theta
      - 3 \, z \, \big[ 1 - (\hat{h} - 1) \, (\gamma - 1) \big] \,
      \cos\theta \, \sin\theta
    }
    ,\nonumber\\\\
    \label{eq:2-018}
    Y_3(\theta) &\equiv&
    \frac{
      - z \, \sin\theta
    }{
      \hat{h} \, \cos\theta + z \, \sin\theta
    },\\
    \label{eq:2-019}
    Z_1(\theta) &\equiv&
    \frac{
      3 \, z^2 \, \sin\theta \, \cos\theta
    }{
      z^2 \, \cos^2\theta + z^2 \, (3 \, \gamma - 4) \, \sin^2\theta
      - 3 \, z \, \big[ 1 - (\hat{h} - 1) \, (\gamma - 1) \big] \,
      \cos\theta \, \sin\theta
    },\nonumber\\\\
    \label{eq:2-020}
    Z_2(\theta) &\equiv&
    \frac{
      z \, \cos\theta
    }{
      \hat{h} \, \cos\theta + z \, \sin\theta
    },
  \end{eqnarray}
  respectively.
  It is noted that
  $\tilde{\Pi}_p$, $\tilde{J}^\mu_p$, and $\tilde{\pi}^{\mu\nu}_p$
  belong to the Q${}_0$ space.
  $\Pi_p$, $J^\mu_p$, and $\pi^{\mu\nu}_p$ are independent of $\theta$,
  which are identically the \textit{microscopic representations of dissipative currents}
  in the literature \cite{mic001}.
  The definitions of
  $\hat{h}$, $\gamma$, and $\Delta^{\mu\nu\rho\sigma}$
  used in Eq.'s (\ref{eq:2-013})-(\ref{eq:2-020}) are
  \begin{eqnarray}
    \label{eq:2-021}
    \hat{h} &\equiv& \frac{e + p}{n\,T},\\
    \label{eq:2-022}
    \gamma &\equiv& 1 + (z^2 - \hat{h}^2 + 5\,\hat{h} - 1)^{-1},\\
    \label{eq:2-023}
    \Delta^{\mu\nu\rho\sigma} &\equiv& \frac{1}{2} \,
    \Big(
    \Delta^{\mu\rho}\,\Delta^{\nu\sigma}
    + \Delta^{\mu\sigma}\,\Delta^{\nu\rho}
    - \frac{2}{3} \, \Delta^{\mu\nu}\,\Delta^{\rho\sigma}
    \Big),
  \end{eqnarray}
  where $\hat{h}$ and $\gamma$ denote
  the reduced enthalpy per particle
  and the ratio of the heat capacities,
  respectively.
  
  With use of $\varphi^{\mu\alpha}_{1p}$ in Eq.(\ref{eq:2-011}),
  we can write down
  $\eta^{\mu\alpha\nu\beta}_1
  = \langle\,\varphi^{\mu\alpha}_1\,,\,L^{-1}\,\varphi^{\nu\beta}_1\,\rangle$
  as
  \begin{eqnarray}
    \label{eq:2-024}
    \eta^{\mu\rho\nu\sigma}_1
    &=&
    - T\,\zeta
    \,
    \Big(Y_1(\theta)\,u^\mu\,u^\rho - Y_2(\theta)\,\Delta^{\mu\rho}\Big)\,
    \Big(Y_1(\theta)\,u^\nu\,u^\sigma - Y_2(\theta)\,\Delta^{\nu\sigma}\Big)
    \nonumber\\
    &&{}+
    T^2\,\lambda
    \, Y_3^2(\theta) \, (
      u^\mu \, u^\nu \, \Delta^{\rho\sigma}
    + u^\mu \, u^\sigma \, \Delta^{\rho\nu}
    + u^\rho \, u^\nu \, \Delta^{\mu\sigma}
    + u^\rho \, u^\sigma \, \Delta^{\mu\nu}
    )\nonumber\\
    &&{}- 2\,T\,\eta
    \, \Delta^{\mu\rho\nu\sigma},\\
    \label{eq:2-025}
    \eta^{\mu\rho\nu 4}_1 = \eta^{\nu 4\mu\rho}_1
    &=&
    - T\,\zeta
    \,
    \Big(
    Y_1(\theta) \, u^\mu \, u^\rho - Y_2(\theta) \, \Delta^{\mu\rho}
    \Big)
    \, Z_1(\theta) \, u^\nu\nonumber\\
    &&{}+
    T^2\,\lambda
    \, Y_3(\theta) \, Z_2(\theta) \, ( u^\mu \, \Delta^{\rho\nu} +
    u^\rho \, \Delta^{\mu\nu} ),\\
    \label{eq:2-026}
    \eta^{\mu 4\nu 4}_1
    &=&
    -T\,\zeta
    \, Z^2_1(\theta) \, u^\mu \, u^\nu
    +
    T^2\,\lambda
    \, Z^2_2(\theta) \, \Delta^{\mu\nu}.
  \end{eqnarray}
  Here, we have introduced the transport coefficients, i.e., the bulk viscosity $\zeta$,
  the heat conductivity $\lambda$,
  and the shear viscosity $\eta$ given by
  \begin{eqnarray}
    \label{eq:2-027}
    \zeta &\equiv& - \frac{1}{T} \, \langle \, \tilde{\Pi}\,,\,L^{-1}\,\tilde{\Pi} \,
    \rangle,\\
    \label{eq:2-028}
    \lambda &\equiv& \frac{1}{3\,T^2} \, \langle \, \tilde{J}^\mu\,,\,L^{-1}\,\tilde{J}_\mu \,
    \rangle,\\
    \label{eq:2-029}
    \eta &\equiv& - \frac{1}{10\,T} \, \langle \,
    \tilde{\pi}^{\mu\nu}\,,\,
    L^{-1}\,\tilde{\pi}_{\mu\nu} \,
    \rangle,
  \end{eqnarray}
  respectively.
  In \S\ref{sec:4}, we show that
  these quantities are independent of $\theta$.
  In the derivations of Eq.'s (\ref{eq:2-024})-(\ref{eq:2-026}),
  we have used the following identities
  \begin{eqnarray}
    \label{eq:2-030}
    \langle \, \tilde{J}^\mu\,,\,L^{-1}\,\tilde{J}^\nu \, \rangle
    &=& \frac{1}{3} \, \langle \, \tilde{J}^a\,,\,L^{-1}\,\tilde{J}_a \,
    \rangle \, \Delta^{\mu\nu},\\
    \label{eq:2-031}
    \langle \, \tilde{\pi}^{\mu\nu}\,,\,L^{-1}\,\tilde{\pi}^{\rho\sigma} \, \rangle
    &=& \frac{1}{5} \, \langle \, \tilde{\pi}^{ab}\,,\,L^{-1}\,\tilde{\pi}_{ab} \,
    \rangle \, \Delta^{\mu\nu\rho\sigma}.
  \end{eqnarray}
  
  Now let us rewrite the expressions of
  $\zeta$, $\lambda$, and $\eta$ in a more familiar form, i.e.,
  the Green-Kubo formula \cite{green-kubo} 
  in the linear response theory \cite{bbgky,lin-res}.
  With use of the identity
  \begin{eqnarray}
    \label{eq:2-032}
    \sum_q \, \big[ L^{-1} \big]_{pq}\, (\tilde{\Pi}_q,\,\tilde{J}^\mu_q,\,\tilde{\pi}^{\mu\nu}_q)
    = - \int_0^\infty\!\!\mathrm{d}s\,\,\,\sum_q \, \big[ \mathrm{e}^{sL} \big]_{pq}\,
    (\tilde{\Pi}_q,\,\tilde{J}^\mu_q,\,\tilde{\pi}^{\mu\nu}_q),
  \end{eqnarray}
  we can rewrite Eq's (\ref{eq:2-027})-(\ref{eq:2-029}) as
  \begin{eqnarray}
    \label{eq:2-033}
    \zeta &=& \int_0^\infty\!\!\mathrm{d}s\,\,\,R_\zeta(s),\\
    \label{eq:2-034}
    \lambda &=& \int_0^\infty\!\!\mathrm{d}s\,\,\,R_\lambda(s),\\
    \label{eq:2-035}
    \eta &=& \int_0^\infty\!\!\mathrm{d}s\,\,\,R_\eta(s),
  \end{eqnarray}
  where
  \begin{eqnarray}
    \label{eq:2-036}
    R_\zeta(s) &\equiv& \frac{1}{T} \, \langle \, \tilde{\Pi}(0)\,,\,\tilde{\Pi}(s) \,
    \rangle,\\
    \label{eq:2-037}
    R_\lambda(s) &\equiv& - \frac{1}{3\,T^2} \, \langle
    \, \tilde{J}^\mu(0)\,,\,\tilde{J}_\mu(s) \, \rangle,\\
    \label{eq:2-038}
    R_\eta(s) &\equiv& \frac{1}{10\,T} \, \langle
    \, \tilde{\pi}^{\mu\nu}(0)\,,\,\tilde{\pi}_{\mu\nu}(s) \,\rangle,
  \end{eqnarray}
  with
  \begin{eqnarray}
    \label{eq:2-039}
    (\tilde{\Pi}_p(s),\,\tilde{J}^\mu_p(s),\,\tilde{\pi}^{\mu\nu}_p(s))
    \equiv
    \sum_q \, \big[ \mathrm{e}^{sL} \big]_{pq}\,
    (\tilde{\Pi}_q,\,\tilde{J}^\mu_q,\,\tilde{\pi}^{\mu\nu}_q).
  \end{eqnarray}
  It is noted that $R_\zeta(s)$, $R_\lambda(s)$, and $R_\eta(s)$
  in Eq.'s (\ref{eq:2-036})-(\ref{eq:2-038}) are called
  the \textit{relaxation function} in the linear response theory \cite{lin-res,bbgky}.
  
  Finally,
  the dissipative currents $\delta J^{\mu\alpha}_\mathrm{1st}$
  are obtained from $\eta^{\mu\alpha\nu\beta}_1$ in Eq.'s
  (\ref{eq:2-024})-(\ref{eq:2-026})
  and $\bar{X}_{\nu\beta}$ in Eq.(\ref{eq:2-010}), as follows:
  \begin{eqnarray}
    \label{eq:2-040}
    \delta J^{\mu\alpha}_\mathrm{1st}
    =\left\{
    \begin{array}{ll}
      \displaystyle{
        \zeta\,\Big( Y_1(\theta)\,u^\mu\,u^\nu
        - Y_2(\theta)\,\Delta^{\mu\nu} \Big) \, X_{\Pi}
      }
      & \displaystyle{} \\[2mm]
      \displaystyle{
        \hspace{2cm}{}- T \, \lambda \, Y_3(\theta) \, (u^\mu \, X_J^\nu + u^\nu \, X_J^\mu)
        + 2\,\eta\,X_\pi^{\mu\nu} = \delta T^{\mu\nu}
      }
      & \displaystyle{\mathrm{for}\,\,\,\alpha = \nu,} \\[2mm]
      \displaystyle{
        \zeta\,Z_1(\theta)\,u^\mu\,X_\Pi - T\,\lambda\,Z_2(\theta)
        \, X_J^\mu  = m\,\delta N^{\mu}
      }
      & \displaystyle{\mathrm{for}\,\,\,\alpha = 4,}
    \end{array}
    \right.\nonumber\\
  \end{eqnarray}
  where the new thermodynamic forces
  $X_\Pi$, $X_J^\mu$, and $X_\pi^{\mu\nu}$ are defined by
  \begin{eqnarray}
    \label{eq:2-041}
    X_\Pi &\equiv&
    - T \, \Big( Y_1(\theta)\,u^\mu\,u^\nu - Y_2(\theta)\,\Delta^{\mu\nu}
    \Big)\,\bar{X}_{\mu\nu}
    - T \, Z_1(\theta)\,u^\mu\,\bar{X}_{\mu 4}\nonumber\\
    &=& - Y_2(\theta) \, \nabla\cdot u,\\
    \label{eq:2-042}
    X_J^\mu &\equiv& - T \, \Delta^{\mu\nu}\,\Big(
    Y_3(\theta)\,u^\rho( \bar{X}_{\nu\rho} + \bar{X}_{\rho\nu})
    + Z_2(\theta)\,\bar{X}_{\nu 4}
    \Big)\nonumber\\
    &=& Y_3(\theta)\,T\,\nabla_\mu(1/T) - Z_2(\theta)\,z^{-1}\,\nabla_\mu(\mu/T),\\
    \label{eq:2-043}
    X_\pi^{\mu\nu} &\equiv& - T \,
    \Delta^{\mu\nu\rho\sigma}\,\bar{X}_{\rho\sigma}\nonumber\\
    &=&
    \Delta^{\mu\nu\rho\sigma}\,\nabla_\rho u_\sigma.
  \end{eqnarray}
  Here, the following relations have been used:
  $u^{\mu} \, \bar{X}_{\mu\alpha} = 0$ and $u^{\nu} \, \nabla_{\mu}
  u_{\nu} = 0$.
  Thus, we have arrived at the generic form of the currents
  $J^{\mu\alpha}_{\mathrm{1st}} = J^{(0)\mu\alpha}_\mathrm{1st}
  + \delta J^{\mu\alpha}_\mathrm{1st}$
  for an arbitrary $\theta$ specifying the local rest frame of the flow.
  Equations (\ref{eq:2-001}) and (\ref{eq:2-040}) together with
  Eq.'s (\ref{eq:2-041})-(\ref{eq:2-043})
  are the main results in this section.
  
  To have a physical intuition into
  the effects of the dissipation on the hydrodynamic equations,
  we also decompose $\delta J^{\mu\alpha}_\mathrm{1st}$ in
  Eq.(\ref{eq:2-040})
  into various tensors
  \begin{eqnarray}
    \label{eq:2-044}
    \delta J^{\mu\alpha}_\mathrm{1st} = \left\{
    \begin{array}{ll}
      \displaystyle{
        \delta e \, u^\mu \, u^\nu - \delta p \, \Delta^{\mu\nu} + Q^\mu
        \, u^\nu + Q^\nu \, u^\mu + \Pi^{\mu\nu} = \delta T^{\mu\nu}
      }
      & \displaystyle{\mathrm{for}\,\,\,\alpha = \nu,} \\[2mm]
      \displaystyle{
        m\,\delta n \, u^\mu + m\,\nu^\mu =  m \, \delta N^{\mu},
      }
      & \displaystyle{\mathrm{for}\,\,\,\alpha = 4,}
    \end{array}
    \right.
  \end{eqnarray}
  where
  \begin{eqnarray}
    \label{eq:2-045}
    \delta e &\equiv& \delta J^{\mu\nu}_\mathrm{1st} \, u_\mu \, u_\nu
    = \delta T^{\mu\nu} \, u_\mu \, u_\nu ,\\
    \label{eq:2-046}
    \delta p &\equiv& -1/3 \, \delta J^{\mu\nu}_\mathrm{1st} \,
    \Delta_{\mu\nu}
    = -1/3 \, \delta T^{\mu\nu} \, \Delta_{\mu\nu},\\
    \label{eq:2-047}
    Q^\mu &\equiv& \delta J^{\nu\rho}_\mathrm{1st} \, u_\nu
    \Delta_\rho^{\,\,\,\mu}
    =\delta T^{\nu\rho} \, u_\nu \, \Delta_\rho^{\,\,\,\mu},\\
    \label{eq:2-048}
    \Pi^{\mu\nu} &\equiv& \delta J^{\rho\sigma}_\mathrm{1st} \,
    \Delta_{\rho\sigma}^{\,\,\,\,\,\,\mu\nu}
    = \delta T^{\rho\sigma} \, \Delta_{\rho\sigma}^{\,\,\,\,\,\,\mu\nu},\\
    \label{eq:2-049}
    \delta n &\equiv& m^{-1}\delta J^{\mu 4}_\mathrm{1st} \, u_\mu
    = \delta N^{\mu} \, u_\mu,\\
    \label{eq:2-050}
    \nu^\mu &\equiv& m^{-1}\delta J^{\nu 4}_\mathrm{1st}\,
    \Delta_\nu^{\,\,\,\mu}
    = \delta N^{\nu } \, \Delta_\nu^{\,\,\,\mu}.
  \end{eqnarray}
  Here, $\delta e$, $\delta p$, and $\delta n$ represent the
  contribution of the dissipation
  to the internal energy, pressure and particle-number density, respectively.
  We stress that the dissipation in the relativistic system
  generically gives rise to an additional contribution
  to such would-be thermodynamic quantities,
  as well as the dissipative currents, $Q^{\mu}$, $\Pi^{\mu\nu}$, and $\nu^{\mu}$.
  Furthermore,
  it is to be noted that the existence of $\delta e$ defined in Eq.(\ref{eq:2-045})
  is directly related with the definition of the local rest frame
  as is seen from Eq.(\ref{eq:ansatz-i}).
  
  Inserting Eq.(\ref{eq:2-040}) into Eq.'s (\ref{eq:2-045})-(\ref{eq:2-050}),
  we have
  \begin{eqnarray}
    \label{eq:2-051}
    \delta e &=& \zeta \,
    Y_1(\theta) \, X_\Pi\nonumber\\
    &=& - \zeta \,
    Y_1(\theta) \, Y_2(\theta) \, \nabla\cdot u,\\
    \label{eq:2-052}
    \delta p &=& \zeta \,
    Y^2(\theta) \, X_\Pi\nonumber\\
    &=& - \zeta \,
    Y^2_2(\theta) \, \nabla\cdot u,\\
    \label{eq:2-053}
    Q^\mu &=&  - T \, \lambda \, Y_3(\theta) \, X_J^\mu\nonumber\\
    &=& - T \, \lambda \, \Big(
    Y_3^2(\theta) \, T \, \nabla^\mu(1/T)
    -
    Y_3(\theta) \, Z_2(\theta) \, z^{-1} \, \nabla^\mu(\mu/T)
    \Big),\\
    \label{eq:2-054}
    \Pi^{\mu\nu} &=& 2 \, \eta \,
    X_\pi^{\mu\nu}\nonumber\\
    &=& 2 \, \eta \,
    \Delta^{\mu\nu\rho\sigma} \, \nabla_\rho u_\sigma,\\
    \label{eq:2-055}
    m\,\delta n &=& \zeta \,
    Z_1(\theta) \, X_\Pi\nonumber\\
    &=& - \zeta \,
    Y_2(\theta) \, Z_1(\theta) \, \nabla\cdot u,\\
    \label{eq:2-056}
    m\,\nu^\mu &=& - T \, \lambda \, Z_2(\theta) \, X_J^\mu\nonumber\\
    &=& - T \, \lambda \, \Big(
    Y_3(\theta) \, Z_2(\theta) \, T \, \nabla^\mu(1/T)
    -
    Z^2_2(\theta) \, z^{-1} \, \nabla^\mu(\mu/T)
    \Big).
  \end{eqnarray}
  Equations (\ref{eq:2-051})-(\ref{eq:2-056}) also constitute main results
  in this section,
  and we call them the \textit{constitutive equations}.
  
  We shall consider here
  the constitutive equations
  only for a few values of $\theta$ which satisfies Eq.(\ref{eq:2-007}),
  but show that
  our generic form of the constitutive equations
  contains
  ones in the two well-known frames,
  i.e., the energy one and the particle one,
  which were introduced in \S\ref{sec:1}.
  
  \subsection{
    Energy frame
  }
  With the simplest choice $\theta=0$, i.e., $\Ren{a}^\mu_p = u^\mu$,
  we have
  \begin{eqnarray}
    \label{eq:2-057}
    \delta e &=& 0,\\
    \label{eq:2-058}
    \delta p &=& - \zeta \, \nabla\cdot u,\\
    \label{eq:2-059}
    Q^\mu &=& 0,\\
    \label{eq:2-060}
    \Pi^{\mu\nu} &=& 2 \, \eta \,
    \Delta^{\mu\nu\rho\sigma} \, \nabla_\rho u_\sigma,\\
    \label{eq:2-061}
    m\,\delta n &=& 0,\\
    \label{eq:2-062}
    m\,\nu^\mu &=& \lambda \, \frac{m}{\hat{h}^2} \, \nabla^\mu\frac{\mu}{T},
  \end{eqnarray}
  where we have used
  the following relations obtained from Eq.'s (\ref{eq:2-016})-(\ref{eq:2-020}),
  \begin{eqnarray}
    \label{eq:2-063}
    Y_1(0) &=& 0,\\
    \label{eq:2-064}
    Y_2(0) &=& 1,\\
    \label{eq:2-065}
    Y_3(0) &=& 0,\\
    \label{eq:2-066}
    Z_1(0) &=& 0,\\
    \label{eq:2-067}
    Z_2(0) &=& z/\hat{h}.
  \end{eqnarray}
  It is noted that
  the choice $\theta = 0$ gives $Q^\mu = 0$ as well as $\nu^\mu \neq 0$,
  which tells us that
  our hydrodynamic equation with $\theta = 0$ is the one in the energy frame.
  We remark that the dissipation gives rise to
  an additional pressure $\delta p$
  while the dissipative internal energy $\delta e$ and 
  particle-number density $\delta n$ are absent in this frame.
  Thus, the conditions of fit (\ref{eq:cond-fit-n-2})
  used in the usual Chapman-Enskog expansion method
  is found to be compatible with the underlying Boltzmann equation.
  
  Using Eq.'s (\ref{eq:2-001}) and (\ref{eq:2-044}),
  we can rewrite
  $J^{\mu\alpha}_{\mathrm{1st}} =
  J^{(0)\mu\alpha}_{\mathrm{1st}} + \delta J^{\mu\alpha}_{\mathrm{1st}}$ as
  \begin{eqnarray}
    \label{eq:2-068}
    T^{\mu\nu} &=& e\,u^\mu\,u^\nu - (p - \zeta \, \nabla\cdot u)\,\Delta^{\mu\nu}
    + 2 \, \eta \, \Delta^{\mu\nu\rho\sigma} \, \nabla_\rho u_\sigma,\\
    \label{eq:2-069}
    N^\mu &=& n\,u^\mu + \lambda \,
    \frac{1}{\hat{h}^2} \, \nabla^\mu\frac{\mu}{T}.
  \end{eqnarray}
  We should notice that
  $T^{\mu\nu}$
  and $N^\mu$ in Eq.'s (\ref{eq:2-068}) and (\ref{eq:2-069})
  completely agree with
  the energy-frame currents proposed by Landau and Lifshitz \cite{hen002}.
  Indeed,
  the respective dissipative parts
  $\delta T^{\mu\nu}$ and $\delta N^{\mu}$
  in Eq.'s (\ref{eq:2-068}) and (\ref{eq:2-069}) meet Landau-Lifshitz's constraints
  given by Eq.'s (\ref{eq:ansatz-i}), (\ref{eq:ansatz-ii}), and (\ref{eq:ansatz-iv}).
  This was actually anticipated:
  In fact, if we take the simplest choice $\Ren{a}_p^\mu = u^\mu$,
  Eq.(\ref{eq:quantity}) tells us that
  the physical quantity transported by each particle
  in the system
  reads
  \begin{eqnarray}
    p \cdot u.
  \end{eqnarray}
  To make clearer the physical meaning of $(p \cdot u)$,
  we take the non-relativistic limit of this quantity:
  \begin{eqnarray}
    \label{eq:2-070}
    p \cdot u \sim m + \frac{m}{2} \, \Big|\frac{\Vec{p}}{m} - \Vec{u}\Big|^2,
  \end{eqnarray}
  where $u^\mu = (u^0,\,\Vec{u})$ and $p^\mu = (p^0,\,\Vec{p})$.
  Equation (\ref{eq:2-070}) shows that
  $(p \cdot u)$ can be identified as
  the kinetic energy of the fluid component
  measured in the rest frame of $u^\mu$.
  Thus, it is natural that
  the currents $J^{\mu\alpha}_{\mathrm{1st}}$
  with the choice  $\Ren{a}_p^\mu = u^\mu$ becomes ones
  in the energy frame adopted by Landau and Lifshitz.
  
  \subsection{
    Particle frame
  }
  Another simple choice $\theta=\pi/2$, i.e.,
  $\Ren{a}^\mu_p = m/(p\cdot u) \, u^\mu$,
  gives the following constitutive equations
  \begin{eqnarray}
    \label{eq:2-071}
    \delta e &=& - 3 \, \zeta \, (3 \, \gamma - 4)^{-2} \, \nabla\cdot u,\\
    \label{eq:2-072}
    \delta p &=& - \zeta \, (3 \, \gamma - 4)^{-2} \, \nabla\cdot u,\\
    \label{eq:2-073}
    Q^\mu &=& \lambda \, \nabla^\mu T,\\
    \label{eq:2-074}
    \Pi^{\mu\nu} &=& 2 \, \eta \,
    \Delta^{\mu\nu\rho\sigma} \, \nabla_\rho u_\sigma,\\
    \label{eq:2-075}
    m\,\delta n &=& 0,\\
    \label{eq:2-076}
    m\,\nu^\mu &=& 0.
  \end{eqnarray}
  We have utilized the following relations
  \begin{eqnarray}
    \label{eq:2-077}
    Y_1(\pi/2) &=& -3 \, (3 \, \gamma - 4)^{-1},\\
    \label{eq:2-078}
    Y_2(\pi/2) &=& - (3 \, \gamma - 4)^{-1},\\
    \label{eq:2-079}
    Y_3(\pi/2) &=& 1,\\
    \label{eq:2-080}
    Z_1(\pi/2) &=& 0,\\
    \label{eq:2-081}
    Z_2(\pi/2) &=& 0,
  \end{eqnarray}
  which are derived from Eq.'s (\ref{eq:2-016})-(\ref{eq:2-020}), respectively.
  We see that $\nu^\mu = 0$ while $Q^\mu \neq 0$,
  which tells us that
  our generic equation with $\theta = \pi/2$ becomes the hydrodynamic equation
  in the particle frame or Eckart frame.
  We also note that
  the dissipative internal energy $\delta e$
  and pressure $\delta p$ are finite in this frame in our formalism
  while $\delta n = 0$;
  we remark that the presence of $\delta p$
  was also the case in the energy frame.
  
  In terms of $T^{\mu\nu}$ and $N^\mu$, we have
  \begin{eqnarray}
    \label{eq:2-082}
    T^{\mu\nu} &=& \Big(e - 3\,\zeta \, (3 \, \gamma - 4)^{-2} \,\nabla\cdot u\Big)\,u^\mu\,u^\nu
    - \Big(p - \zeta \, (3 \, \gamma - 4)^{-2} \,\nabla\cdot u\Big)\,\Delta^{\mu\nu}\nonumber\\
    &&{}+ \lambda \,  (u^\mu \, \nabla^\nu T + u^\nu \, \nabla^\mu T)
    + 2 \, \eta \, \Delta^{\mu\nu\rho\sigma} \, \nabla_\rho u_\sigma,\\
    \label{eq:2-083}
    N^\mu &=& n\,u^\mu.
  \end{eqnarray}
  It is noteworthy that
  the set of Eq.'s (\ref{eq:2-082}) and (\ref{eq:2-083})
  is different from the phenomenological equations
  in the particle frame by Eckart,
  but may be regarded as a corrected version of that by Stewart \cite{mic003},
  which is derived from the relativistic Boltzmann equation
  on the basis of the moment method:
  Indeed,
  the respective dissipative parts
  $\delta T^{\mu\nu}$ and $\delta N^{\mu}$
  in Eq.'s (\ref{eq:2-082}) and (\ref{eq:2-083}) satisfy the conditions
  (\ref{eq:ansatz-v}), (\ref{eq:ansatz-ii}), and (\ref{eq:ansatz-iii})
  as Stewart's equation does,
  but not Eckart's constraints
  given by Eq.'s (\ref{eq:ansatz-i}), (\ref{eq:ansatz-ii}), and (\ref{eq:ansatz-iii}).
  The latter also implies that the conditions of fit (\ref{eq:cond-fit-e-2}) imposed
  in the usual Chapman-Enskog expansion method is not compatible with 
  the underlying Boltzmann equation.
  
  Now it is interesting that
  the above constraints on the energy-momentum tensor
  do not necessarily determine the detailed structure of it.
  In fact, Eq.(\ref{eq:2-082}) is different from that of Stewart,
  which is obtained from our equation
  through the replacements
  \begin{eqnarray}
    -\zeta \, ( 3 \, \gamma - 4)^{-2} \, \nabla \cdot u &\rightarrow&
    + \zeta \, ( 3 \, \gamma - 4)^{-1} \, \nabla \cdot u,\\
    \lambda \, \nabla^{\mu} T &\rightarrow& \lambda \, (\nabla^{\mu} T -
    T \, Du^{\mu}),
  \end{eqnarray}
  with $D \equiv u^{\mu} \, \partial_{\mu}$,
  although both of them satisfy the constraints
  (\ref{eq:ansatz-v}), (\ref{eq:ansatz-ii}), and (\ref{eq:ansatz-iii}).
  There are two noteworthy points in Eq.(\ref{eq:2-082}):
  First,
  the sign of the thermodynamic force
  with the bulk viscosity in our equation
  is the same as that in the Landau-Lifshitz equation
  and opposite that in the Stewart equation,
  because $\gamma > 4/3$ \cite{mic001}.
  Second,
  the heat flow $\Delta_{\mu\nu} \, u_\rho \,  \delta T^{\nu\rho} = Q^\mu$
  does not contain the time-derivative term, $Du_\mu$, i.e.,
  the acceleration term,
  which is often interpreted as
  one of the important relativistic effects \cite{hen001}.
  
  Now
  a natural question is which equation is preferable,
  ours or Stewart's or Eckart's?
  It is known
  that there have been no hydrodynamic equation
  including Eckart's and Stewart's
  which has the stable thermal equilibrium state,
  as is shown by Lindblum and Hiskock \cite{hyd002}.
  Owing to the absence of $Du^\mu$,
  we can show that the thermal equilibrium solution of our equation becomes stable
  against a small perturbation \cite{env010},
  in contrast to the Eckart and the original Stewart equations.
  Thus, we claim that
  our equation (\ref{eq:2-082}) is the proper hydrodynamic equation
  free from the instability problem
  in the particle frame.
  We emphasize that such an equation has been obtained
  for the first time, as far as we are aware of.
  A detailed and general account of this stability issue
  will be given in \S\ref{sec:6}.
  
  \subsection{
    A new frame
  }
  The other interesting choice is $\theta = \pi/4$,
  for which we have
  $\Ren{a}_p^\mu = ((p \cdot u) + m)/(\sqrt{2} \, p \cdot u) \, u^\mu$,
  and
  \begin{eqnarray}
    \label{eq:2-084}
    \delta e &=& 0,\\
    \label{eq:2-085}
    \delta p &=& 0,\\
    \label{eq:2-086}
    Q^\mu &=& \lambda \, \Big(\frac{z}{\hat{h}+z}\Big)^2 \,
    \Big(\nabla^\mu T - z^{-1} \, T \, \nabla^\mu \frac{\mu}{T}\Big),\\
    \label{eq:2-087}
    \Pi^{\mu\nu} &=& 2 \, \eta \,
    \Delta^{\mu\nu\rho\sigma} \, \nabla_\rho u_\sigma,\\
    \label{eq:2-088}
    m\,\delta n &=& 0,\\
    \label{eq:2-089}
    m\,\nu^\mu &=& - \lambda \, \Big(\frac{z}{\hat{h}+z}\Big)^2 \,
    \Big(\nabla^\mu T - z^{-1} \, T \, \nabla^\mu \frac{\mu}{T}\Big).
  \end{eqnarray}
  Here, we have used the following relations obtained from
  Eq.'s (\ref{eq:2-016})-(\ref{eq:2-020}),
  \begin{eqnarray}
    \label{eq:2-090}
    Y_1(\pi/4) &=& -\frac{3\,z^2}{z^2 + z^2 \, (3\,\gamma-4) -
      3\,z\,\big[ 1 - (\hat{h} - 1)\,(\gamma - 1) \big]},\\
    \label{eq:2-091}
    Y_2(\pi/4) &=& 0,\\
    \label{eq:2-092}
    Y_3(\pi/4) &=& -\frac{z}{\hat{h} + z},\\
    \label{eq:2-093}
    Z_1(\pi/4) &=&  \frac{3\,z^2}{z^2 + z^2 \, (3\,\gamma-4) -
      3\,z\,\big[ 1 - (\hat{h} - 1)\,(\gamma - 1) \big]},\\
    \label{eq:2-094}
    Z_2(\pi/4) &=& \frac{z}{\hat{h} + z}.
  \end{eqnarray}
  We note that the present choice $\theta = \pi/4$ gives
  a local rest frame where
  $Q^\mu \neq 0$ and $\nu^\mu \neq 0$;
  recall that $Q^\mu=0$ ($\nu^\mu = 0$) in the energy (particle) frame.
  Thus, our generic hydrodynamic equation with $\theta = \pi/4$ gives the equation
  in a local rest frame
  which is neither the energy frame nor the particle frame.
  We also remark that
  the dissipative internal energy, pressure, and particle-number density
  are absent in this frame;
  $\delta e=\delta p=\delta n =0$.
  
  The alternative expressions in terms of $T^{\mu\nu}$ and $N^\mu$ read
  \begin{eqnarray}
    \label{eq:2-095}
    T^{\mu\nu} &=& e\,u^\mu\,u^\nu
    - p\,\Delta^{\mu\nu}\nonumber\\
    &&{}+
    \lambda \, \Big(\frac{z}{\hat{h}+z}\Big)^2 \,\Bigg[
    u^\mu \,
    \Big(\nabla^\nu T - z^{-1} \, T \, \nabla^\nu \frac{\mu}{T}\Big)+
    u^\nu \,
    \Big(\nabla^\mu T - z^{-1} \, T \, \nabla^\mu \frac{\mu}{T}\Big)\Bigg]\nonumber\\
    &&{}+ 2 \, \eta \, \Delta^{\mu\nu\rho\sigma} \, \nabla_\rho u_\sigma,\\
    \label{eq:2-096}
    N^\mu &=& n\,u^\mu - \lambda \, \Big(\frac{z}{\hat{h}+z}\Big)^2 \,
    \Big(\nabla^\mu T - z^{-1} \, T \, \nabla^\mu \frac{\mu}{T}\Big).
  \end{eqnarray}
  We see another unique feature of this frame that
  the bulk pressure term is absent,
  and hence $\zeta$ does not play any role in this frame.
  As far as we know,
  this type of the relativistic dissipative hydrodynamics
  is written down for the first time,
  which was made possible
  by the introduction of the macroscopic-frame vector in the powerful RG method
  \footnote{
    In a previous short communication \cite{env009},
    we discussed the equation obtained with an erroneous setting $\theta = -\pi/4$.
    We note that this choice is not appropriate
    and the resulting equation should be abandoned
    since $-\pi/4$ is out of range of
    $\theta$ (\ref{eq:2-007})
    that guarantees the positive definiteness of the inner product.
  }.
  
  \setcounter{equation}{0}
  \section{
    Discussions
  }
  \label{sec:4-0}
  In this section,
  we shall examine the properties
  of the derived equation in the generic frame
  and
  its advantageous nature over the existing equations in the literature,
  focusing on transport coefficients and local rest frames.
  We give a general proof without recourse to numerical calculations
  that the hydrodynamic equations obtained in our formalism
  have the stable equilibrium state
  on the basis of the positive definiteness of the inner product.
  
  \subsection{
    Frame independence of transport coefficients
  }
  \label{sec:4}
  In this subsection,
  we shall show that the transport coefficients $\zeta$, $\lambda$,
  and $\eta$ in Eq.'s (\ref{eq:2-027})-(\ref{eq:2-029}) obtained
  in our formalism are independent of $\theta$.
  For this purpose,
  we introduce a new linearized collision operator
  \begin{eqnarray}
    \label{eq:3-001}
    \mathcal{L}_{pq} &\equiv& (p\cdot\theta_p) \, L_{pq}\nonumber\\
    &=& - \frac{1}{2!} \, \sum_{p_1} \,
    \frac{1}{p_1^0} \, \sum_{p_2} \, \frac{1}{p_2^0} \, \sum_{p_3} \,
    \frac{1}{p_3^0} \, \omega(p \,,\, p_1|p_2 \,,\, p_3) \,
    f^{\mathrm{eq}}_{p_1} \, ( \delta_{p q} + \delta_{p_1 q} -
    \delta_{p_2 q} - \delta_{p_3 q} ),\nonumber\\
  \end{eqnarray}
  which is independent of $\theta$.
  We note that the inverse matrix of $\mathcal{L}_{pq}$ reads
  \begin{eqnarray}
    \label{eq:3-002}
    \mathcal{L}^{-1}_{pq} = L^{-1}_{pq} \, (q\cdot\theta_q)^{-1}.
  \end{eqnarray}
  Using $\mathcal{L}^{-1}_{pq}$ and
  the $\theta$-independent
  microscopic representations of the dissipative currents
  $(\Pi_p,\,J^\mu_p,\,\pi^{\mu\nu}_p)$ in Eq.'s(\ref{eq:2-013})-(\ref{eq:2-015}),
  we can represent the transport coefficients as
  \begin{eqnarray}
    \label{eq:3-003}
    \zeta &=&  - \frac{1}{T} \, {\langle \, \Pi\,,\,\mathcal{L}^{-1}\,\Pi \,
      \rangle}_\mathrm{eq},\\
    \label{eq:3-004}
    \lambda &=& \frac{1}{3\,T^2} \, {\langle \, J^\mu\,,\,\mathcal{L}^{-1}\,J_\mu \,
      \rangle}_\mathrm{eq},\\
    \label{eq:3-005}
    \eta &=& - \frac{1}{10\,T} \, {\langle \, \pi^{\mu\nu}\,,\,\mathcal{L}^{-1}\,\pi_{\mu\nu} \,
      \rangle}_\mathrm{eq},
  \end{eqnarray}
  where we have introduced the new inner product
  \begin{eqnarray}
    \label{eq:3-006}
	  {\langle \, \varphi\,,\,\psi \, \rangle}_\mathrm{eq} \equiv \sum_p \,
	  \frac{1}{p^0} \, f^\mathrm{eq}_p \, \varphi_p
	  \, \psi_p.
  \end{eqnarray}
  It is noted that
  this inner product is identically the thermal average,
  which is independent of $\theta$.
  Equations (\ref{eq:3-003})-(\ref{eq:3-005}) show that
  $\zeta$, $\lambda$, and $\eta$ are $\theta$-invariant,
  and agree with the definitions used in the literature.
  
  \subsection{
    Absence of macroscopic-frame vector leading to Eckart's
    particle-frame equation
  }
  \label{sec:5}
  In \S \ref{sec:3},
  we have seen that
  the hydrodynamic equation in the particle frame
  does not take the form of Eckart,
  if the equation is to be derived from the relativistic Boltzmann equation.
  This fact might not be well known apart from some exceptions \cite{hyd001}.
  For completeness and instructive purpose,
  we shall here give a proof
  that
  there can not exist a macroscopic-frame vector $\Ren{a}^\mu_p$
  that leads to Eckart's constraints,
  (\ref{eq:ansatz-i}), (\ref{eq:ansatz-ii}), and (\ref{eq:ansatz-iii}).
  It will be found in the course of the proof that
  the orthogonality condition between the P${}_0$ and Q${}_0$ space
  exactly corresponds to the phenomenological constraints
  imposed to the dissipative part of the energy-momentum tensor
  $\delta T^{\mu\nu}$ and the particle current $\delta N^\mu$
  defined in
  Eq.'s (\ref{eq:N-dissipative}) and (\ref{eq:T-dissipative}).
  
  We first note that the dissipative currents 
  $\delta J^{\mu\alpha}_\mathrm{1st}$
  given in Eq.(\ref{eq:1-070})
  can be expressed as
  \begin{eqnarray}
    \label{eq:4-004}
    \delta J^{\mu\alpha}_\mathrm{1st} = \sum_p \, \frac{1}{p^0} \, p^\mu \,
    \varphi^\alpha_{0p} \, f^\mathrm{eq}_p \, \bar{\phi}_p,
  \end{eqnarray}
  where $\bar{\phi}_p$ is defined by
  \begin{eqnarray}
    \label{eq:4-003}
    \bar{\phi}_p \equiv - \big[ L^{-1} \, Q_0 \, f^{\mathrm{eq}-1}\, F \big]_p
    = \big[ L^{-1} \, \varphi^{\mu\alpha}_1 \big]_p \, \bar{X}_{\mu\alpha},
  \end{eqnarray}
  where $\varphi^{\mu\alpha}_{1p}$
  are the first-excited modes (\ref{eq:1-077})
  and $\bar{X}_{\mu\alpha}$
  the thermodynamic forces defined by Eq.(\ref{eq:1-073}).
  Note that $\bar{\phi}_p$ belongs to the Q${}_0$ space
  and is accordingly orthogonal to the five vectors
  $\varphi^\alpha_{0p}$
  belonging to the  P${}_0$ space;
  \begin{eqnarray}
    \label{eq:4-005}
    \langle \, \varphi^\alpha_0 \,,\, \bar{\phi}  \, \rangle = 0
    \,\,\,\mathrm{for}\,\,\,\alpha = 0,\,1,\,2,\,3,\,4.
  \end{eqnarray}
  Here,  the inner product is defined by Eq.(\ref{eq:1-034}) where
  $\Ren{a}^\mu_p$ enters in the form of $(p \cdot \Ren{a}_p)$.
  We are going to show that
  these five identical equations
  exactly correspond to the ansatz's on
  $\delta T^{\mu\nu}$ and $\delta N^\mu$;
  see Eq.'s (\ref{eq:N-dissipative}) and (\ref{eq:T-dissipative}).
  
  Firstly,
  let us take the case of the energy frame with the choice (A) $\Ren{a}^\mu_p = u^\mu$.
  Then, Eq.(\ref{eq:4-005}) is reduced to
  \begin{eqnarray}
    \label{eq:4-006}
    \sum_p \, \frac{1}{p^0} \, (p \cdot u) \, f^\mathrm{eq}_p \,
    \varphi^\alpha_{0p} \, \bar{\phi}_p = 0,
  \end{eqnarray}
  which actually expresses a set of equations as follows
  \begin{eqnarray}
    \label{eq:4-007}
    u_\nu \, \delta T^{\mu\nu} &=& 0,\\
    \label{eq:4-008}
    u_\mu \, \delta N^\mu &=& 0.
  \end{eqnarray}
  Note that Eq.(\ref{eq:4-007}) implies the following two equations,
  \begin{eqnarray}
    \label{eq:4-009}
    \delta e = u_\mu \, u_\nu \, \delta T^{\mu\nu} &=& 0,\\
    \label{eq:4-010}
    Q_{\rho} = \Delta_{\rho\mu} \, u_{\nu} \,\delta T^{\mu\nu} &=& 0.
  \end{eqnarray}
  Thus, one can readily see that these equations coincide with
  Landau-Lifshitz's ansatz's,
  (\ref{eq:ansatz-i}), (\ref{eq:ansatz-ii}), and (\ref{eq:ansatz-iv}).
  
  Similarly, with the choice of
  (B) $\Ren{a}^\mu_p = m /(p \cdot u) \, u^\mu$,
  we have
  \begin{eqnarray}
    \label{eq:4-011}
    \sum_p \, \frac{1}{p^0} \, m \, f^\mathrm{eq}_p \,
    \varphi^\alpha_{0p} \, \bar{\phi}_p = 0,
  \end{eqnarray}
  which means that
  \begin{eqnarray}
    \label{eq:4-012}
    \delta N^\mu &=& 0,\\
    \label{eq:4-013}
    \delta T^\mu_{\,\,\,\mu} &=& 0.
  \end{eqnarray}
  Here, we have used the on-shell condition $m^2 = p^\mu \, p_\mu$.
  Equation (\ref{eq:4-012}) is equivalent to the set of equations
  \begin{eqnarray}
    \label{eq:4-014}
    \delta n = u_\mu \, \delta N^\mu &=& 0,\\
    \label{eq:4-015}
    \nu_{\nu} = \Delta_{\nu\mu} \, \delta N^{\mu} &=& 0.
  \end{eqnarray}
  Thus, one sees that
  Stewart's ansatz's,
  (\ref{eq:ansatz-v}), (\ref{eq:ansatz-ii}), and (\ref{eq:ansatz-iii}),
  are derived in the particle frame.
  
  It is now easy to see that
  there exists no macroscopic-frame vector $\Ren{a}^\mu_p$ leading
  to Eckart's ansatz's given by
  Eq.'s (\ref{eq:ansatz-i}), (\ref{eq:ansatz-ii}), and (\ref{eq:ansatz-iii}),
  simultaneously.
  Indeed, in order to lead to the ansatz's (\ref{eq:ansatz-ii}) and (\ref{eq:ansatz-iii})
  on $\delta N^\mu$,
  Eq.(\ref{eq:4-005}) with $\alpha = \mu$ requires that
  $(p \cdot \Ren{a}_p)$ is independent of $p^\mu$, i.e.,
  \begin{eqnarray}
    \label{eq:4-016}
    (p \cdot \Ren{a}_p) = \mathrm{const.}.
  \end{eqnarray}
  On the other hand,
  in order to lead to the ansatz (\ref{eq:ansatz-i})
  on $\delta T^{\mu\nu}$,
  Eq.(\ref{eq:4-005}) with $\alpha = 4$ must lead to
  \begin{eqnarray}
    \label{eq:4-017}
    (p \cdot \Ren{a}_p) = \mathrm{const.} \, \times \, (p \cdot u)^2,
  \end{eqnarray}
  which is in contradiction with Eq.(\ref{eq:4-016}).
  Thus, we conclude that there exists no $\Ren{a}^\mu_p$ leading to
  Eckart's ansatz's (\ref{eq:ansatz-i})-(\ref{eq:ansatz-iii})
  simultaneously.
  
  In short, the phenomenological ansatz's
  on the dissipative parts of
  the energy-momentum tensor $\delta T^{\mu\nu}$
  and the particle current $\delta N^{\mu}$
  have a definite correspondence to the orthogonality conditions
  posed to the dissipative part of the distribution function
  as a solution to the underlying Boltzmann equation.
  In particular,
  whenever the particle frame is taken
  where the particle flow is constructed so as to have no dissipative part,
  the dissipative part of the energy-momentum tensor
  must satisfy Eq.(\ref{eq:ansatz-v})
  but not Eq.(\ref{eq:ansatz-i}).
  This fact implies that the Eckart equation
  can not be the hydrodynamic equation in the particle frame
  as the slow (or infrared) dynamics
  of the relativistic Boltzmann equation,
  and hence nor have microscopic foundation.
  
  \subsection{
    Stability analysis of steady solution of relativistic hydrodynamic
    equation
  }
  \label{sec:6}
  In this subsection, we shall give a general proof
  that steady solutions of
  our generic relativistic dissipative hydrodynamic equation (\ref{eq:1-078})
  are stable against a small perturbation,
  on account of the positive definiteness of
  the inner product (\ref{eq:1-035}).
  Here, a steady solution means that
  it describes a system having a finite homogeneous flow
  with a constant temperature and a constant chemical potential, as follows,
  \begin{eqnarray}
    \label{eq:5-001}
    T(\sigma\,;\,\tau) &=& T_0,\\
    \label{eq:5-002}
    \mu(\sigma\,;\,\tau) &=& \mu_0,\\
    \label{eq:5-003}
    u_\mu(\sigma\,;\,\tau) &=& u_{0\mu},
  \end{eqnarray}
  where $T_0$, $\mu_0$, and $u_{0\mu}$ are constant.
  We note that the steady states include the
  thermal equilibrium state as a special case.
  We remark that such a stability of our equation
  was demonstrated in a previous paper by the present authors \cite{env010}
  by a numerical calculation for a rarefied gas
  with some specific models
  for the cross section in the collision integral.
  The general proof of the stability of our equation
  is given for the first time in the present paper.
  
  To show the stability of the steady solution,
  we apply the so-called linear stability analysis
  to the relativistic dissipative hydrodynamic equation (\ref{eq:1-078}).
    We expand $T$, $\mu$ and $u_\mu$ around the steady solution as follows:
  \begin{eqnarray}
    \label{eq:5-004}
    T(\sigma\,;\,\tau) &=& T_0 + \delta T(\sigma\,;\,\tau),\\
    \label{eq:5-005}
    \mu(\sigma\,;\,\tau) &=& \mu_0 + \delta\mu(\sigma\,;\,\tau),\\
    \label{eq:5-006}
    u_\mu(\sigma\,;\,\tau) &=& u_{0\mu} + \delta u_\mu(\sigma\,;\,\tau).
  \end{eqnarray}
  We assume that
  the higher-order terms than the second order with respect to 
  $\delta T$, $\delta \mu$, and $\delta u_\mu$ can be neglected
  since these quantities are small.
  For convenience,
  we introduce the new variables $\delta X_\alpha$ composed of
  $\delta T$, $\delta \mu$, and $\delta u_\mu$ by
  \begin{eqnarray}
    \label{eq:5-007}
    \delta X_{\alpha} &\equiv& \left\{
    \begin{array}{ll}
      \displaystyle{
        - \delta (u_\mu/T) = - \delta u_\mu/T_0 + \delta
        T \, u_{0\mu}/T^2_0
      }
      & \displaystyle{\mathrm{for}\,\,\,\alpha = \mu}, \\[2mm]
      \displaystyle{
        m^{-1} \, \delta (\mu/T) = m^{-1} \, (\delta\mu/T_0 - \delta T \,
        \mu_0/T^2_0)
      }
      & \displaystyle{\mathrm{for}\,\,\,\alpha = 4}.
    \end{array}
    \right.
  \end{eqnarray}
  Substituting Eq.(\ref{eq:5-007}) into Eq.(\ref{eq:1-078}),
  we obtain the linearized equation governing $\delta X_{\alpha}$ as
  \begin{eqnarray}
    \label{eq:5-008}
    &&\Big( \langle\,\varphi_{0}^\alpha \,,\,
    \varphi^\beta_0 \,\rangle
    + \langle\,\varphi_{0}^\alpha\,,\,
    L^{-1} \, \varphi^{\nu\beta}_1
    \,\rangle \,\Ren{\nabla}_\nu\Big) \,
    \frac{\partial}{\partial \tau}
    \delta X_\beta\nonumber\\
    &&\hspace{2cm}
    {}+
    \Big(
    \langle\,\tilde{\varphi}^{\mu\alpha}_1\,,\,
    \varphi^\beta_0\,\rangle\,\Ren{\nabla}_\mu
    +
    \langle\,\tilde{\varphi}^{\mu\alpha}_1\,,\,
    L^{-1}\,\varphi^{\nu\beta}_1\,\rangle\,\Ren{\nabla}_\mu\,\Ren{\nabla}_\nu
    \Big)\,
    \delta X_\beta
    = 0,
  \end{eqnarray}
  with $\tilde{\varphi}^{\mu\alpha}_{1p} =
  p^\mu\,\varphi^\alpha_{0p}/(p\cdot\Ren{a}_p)$ defined in Eq.(\ref{eq:1-071}).
  Here, we have used the following simple relations
  \begin{eqnarray}
    \label{eq:5-009}
    \delta (f^{\mathrm{eq}}_p) &=& f^{\mathrm{eq}}_p \,
    \varphi^\alpha_{0p} \, \delta X_\alpha,\\
    \label{eq:5-010}
    \delta (\bar{X}_{\mu\alpha}) &=& \Ren{\nabla}_\mu \delta X_\alpha.
  \end{eqnarray}
  We note that
  all of the coefficients in Eq.(\ref{eq:5-008}) take a value of the steady solution
  $(T,\,\mu,\,u_\mu) = (T_0,\,\mu_0,\,u_{0\mu})$.
  With use of
  the orthogonality condition between the P${}_0$ and Q${}_0$ spaces
  and the definitions of $\eta^{\alpha\beta}_0$ and
  $\eta^{\mu\alpha\nu\beta}_1$
  in Eq.'s (\ref{eq:1-043}) and (\ref{eq:1-076}), respectively,
  we can reduce Eq.(\ref{eq:5-008}) to
  \begin{eqnarray}
    \label{eq:5-011}
    A^{\alpha,\beta}\,\frac{\partial}{\partial\tau}\delta X_{\beta} +
    B^{\alpha,\beta}\,\delta X_{\beta} = 0,
  \end{eqnarray}
  where $A^{\alpha,\beta}$ and $B^{\alpha,\beta}$
  are defined by
  \begin{eqnarray}
    \label{eq:5-012}
    A^{\alpha,\beta} &\equiv& \eta^{\alpha\beta}_0,\\
    \label{eq:5-013}
    B^{\alpha,\beta} &\equiv& \langle\,
    \tilde{\varphi}^{\mu\alpha}_1\,,\,\varphi^\beta_0 \,\rangle \,
    \Ren{\nabla}_\mu + \eta^{\mu\alpha\nu\beta}_1 \, \Ren{\nabla}_\mu \, \Ren{\nabla}_\nu.
  \end{eqnarray}
  
  We convert Eq.(\ref{eq:5-011}) into the algebraic equation,
  using the Fourier and Laplace transformations with respect to
  the spatial variable $\sigma^\mu$
  and the temporal variable $\tau$, respectively:
  By inserting
  \begin{eqnarray}
    \label{eq:5-014}
    \delta X_{\alpha}(\sigma\,;\,\tau)
    = \delta \tilde{X}_{\alpha}(k\,;\,\Lambda) \,
    \mathrm{e}^{ik\cdot\sigma - \Lambda\tau},
  \end{eqnarray}
  into Eq.(\ref{eq:5-011}),
  we have
  \begin{eqnarray}
    \label{eq:5-015}
    ( \Lambda\,A^{\alpha,\beta} - \tilde{B}^{\alpha,\beta} )
    \, \delta \tilde{X}_{\beta}
    = 0,
  \end{eqnarray}
  where $\tilde{B}^{\alpha,\beta}$ is defined by
  \begin{eqnarray}
    \label{eq:5-016}
    \tilde{B}^{\alpha,\beta} \equiv i\,\langle\,
    \tilde{\varphi}^{\mu\alpha}_1\,,\,\varphi^\beta_0 \,\rangle \, k_\mu
    - \eta^{\mu\alpha\nu\beta}_1 \, k_\mu \, k_\nu.
  \end{eqnarray}
  In the rest of this section,
  we use the matrix representation
  when no misunderstanding is expected.
  Since we are interested in a nonvanishing solution $\delta \tilde{X} \not= 0$,
  we impose
  \begin{eqnarray}
    \label{eq:5-017}
    \det ( \Lambda\,A - \tilde{B} ) = 0,
  \end{eqnarray}
  which leads to the dispersion relation
  \begin{eqnarray}
    \label{eq:5-018}
    \Lambda = \Lambda(k).
  \end{eqnarray}
  The stability of the steady solution (\ref{eq:5-001})-(\ref{eq:5-003})
  against a small perturbation
  means that $\delta X$ will not increase under time evolution.
  Therefore,
  our next task is to show that
  the real part of $\Lambda(k)$ is non-negative for any $k^\mu$.
  
  Now we first note that the matrix $A$ is a real symmetric and positive-definite matrix:
  \begin{eqnarray}
    \label{eq:5-019}
    w_{\alpha}\,A^{\alpha,\beta}\,w_{\beta}
    &=& \langle\, w_{\alpha}\, \varphi^{\alpha}_0 \,,\,
    w_{\beta} \, \varphi^{\beta}_0 \,\rangle\nonumber\\
    &=& \langle\, \varphi \,,\, \varphi \,\rangle
    > 0 \,\,\,\mathrm{for}\,\,\,w_{\alpha} \neq 0,
  \end{eqnarray}
  with $\varphi_p \equiv w_{\alpha}\, \varphi^{\alpha}_{0p}$.
  Here,
  we have used the positive definiteness
  of the inner product (\ref{eq:1-035}).
  Equation (\ref{eq:5-019}) means that
  the inverse matrix $A^{-1}$ exists,
  and $A^{-1}$ is also a real symmetric positive-definite matrix.
  Thus, using the Cholesky decomposition,
  we can represent $A^{-1}$ as
  \begin{eqnarray}
    \label{eq:5-020}
    A^{-1} = {}^tU\,U,
  \end{eqnarray}
  where $U$ denotes a real matrix and ${}^tU$ a transposed matrix of $U$.
  Substituting Eq.(\ref{eq:5-020}) into Eq.(\ref{eq:5-017}),
  we have
  \begin{eqnarray}
    \label{eq:5-021}
    \det ( \Lambda - U \, \tilde{B}  \, {}^tU ) = 0.
  \end{eqnarray}
  It is noted that
  $\Lambda(k)$ is an eigen value of $U \, \tilde{B} \, {}^tU$.
  
  We notice the following theorem:
  The real part of the eigen value of a complex matrix $C$
  is non-negative
  when the hermite matrix $\mathrm{Re}(C) \equiv (C + C^\dagger)/2$
  is semi-positive definite.
  Applying this theorem to the present case,
  we find that
  the real part of $\Lambda(k)$ becomes non-negative for any $k^\mu$
  when
  $\mathrm{Re}(U \, \tilde{B} \, {}^tU)$
  is a semi-positive definite matrix.
  In fact, we can show that
  $\mathrm{Re}(U \, \tilde{B} \, {}^tU)$ is semi-positive definite as follows:
  \begin{eqnarray}
    \label{eq:5-022}
    w_{\alpha} \, [ \mathrm{Re}(U \, \tilde{B} \, {}^tU) ]^{\alpha,\beta}
    \, w_{\beta}
    &=&
    w_{\alpha} \, [ U \, \mathrm{Re}(\tilde{B}) \, {}^tU ]^{\alpha,\beta}
    \, w_{\beta}\nonumber\\
    &=& [w \, U]_{\alpha} \,
    [\mathrm{Re}(\tilde{B})]^{\alpha,\beta}
    \, [w \, U]_{\beta}\nonumber\\
    &=& - [w \, U]_{\alpha} \, \eta^{\mu\alpha\nu\beta}_1 \,
    k_\mu \, k_\nu \, [w \, U]_{\beta}\nonumber\\
    &=& - \langle\, k_\mu\,[w \, U]_{\alpha} \, \varphi^{\mu\alpha}_1\,,\,
    L^{-1} \, k_\nu\,[w \, U]_{\beta} \, \varphi^{\nu\beta}_1
    \,\rangle\nonumber\\
    &=& - \langle\, \psi \,,\,
    L^{-1} \, \psi \,\rangle \ge 0
    \,\,\,\mathrm{for}\,\,\,w_{\alpha} \neq 0,
  \end{eqnarray}
  with $\psi_p \equiv k_\mu\,[w \, U]_{\alpha} \, \varphi^{\mu\alpha}_{1p}$.
  Therefore, we conclude that
  the steady solution in Eq.'s (\ref{eq:5-001})-(\ref{eq:5-003}) is
  stable against a small perturbation.
  
  We now see that the relativistic dissipative hydrodynamic equation (\ref{eq:1-078})
  obtained by the RG method has a stable steady solution in a definite way.
  It is noteworthy that
  this property is kept in the several equations derived in the setting of
  $\Ren{a}_p^\mu = \theta_p^\mu$, i.e., the energy-frame equation
  in Eq.'s (\ref{eq:2-068}) and (\ref{eq:2-069}),
  the particle-frame equations
  in Eq.'s (\ref{eq:2-082}) and (\ref{eq:2-083}),
  and so on.
  We stress that this is for the first time
  that the relativistic dissipative hydrodynamic equation in the particle frame
  has been obtained whose steady solution is stable
  \footnote{
    In the previous rapid communication \cite{env010},
    we discussed the stability only of the thermal-equilibrium solution, i.e.,
    $u_{0\mu} = (1,\,0,\,0,\,0)$,
    with use of a rarefied-gas approximation,
    where a differential cross section in the collision integral
    is treated as a constant.
  }.
  
  \setcounter{equation}{0}
  \section{
    Summary and Concluding Remarks
  }
  \label{sec:7}
  In this paper,
  we have given a full and detailed account of the derivation
  of the first-order relativistic dissipative hydrodynamic equations
  as a reduction of dynamics
  from the relativistic Boltzmann equation
  with no heuristic assumptions,
  such as the so-called conditions of fit used in the standard methods.
  This was made possible by
  adopting a powerful reduction theory of dynamics, i.e.,
  the renormalization-group method and
  by introducing the macroscopic-frame vector
  which defines the macroscopic local rest frame;
  thereby we successfully have a coarse-grained and Lorentz-covariant equations
  in a generic frame.
  
  The five hydrodynamic modes are naturally identified with the same number of
  the zero modes of the linearized collision operator.
  The excited modes which are to modify the local equilibrium
  distribution function and give the dissipative terms
  are defined so that they are precisely orthogonal to the zero modes
  with a properly defined inner product for the distribution functions.
  It is worth emphasizing that the dissipative terms
  are constructed without recourse to the so-called conditions of fit
  which are imposed to the dissipative terms in an ad-hoc way in the standard methods.
  In our method, the validity of the conditions of fit is checked
  as a property of the derived equations.
  On the basis of
  the nice properties of the properly defined inner product
  and the very nature of the hydrodynamic modes as the zero modes of the linearized
  collision operator,
  we have shown that the so-called Burnett term
  does not affect the hydrodynamic equations.
  
  Our hydrodynamic equation 
  with the dissipative currents given by Eq.(\ref{eq:2-044}),
  which still contains the macroscopic-frame vector,
  is a master equation for a generic local rest frame,
  and derives
  hydrodynamic equations in various local rest frames
  with a specific choice of the macroscopic-frame vector.
  We have shown that our
  energy-momentum tensor and particle current
  in the energy frame,
  given by Eq.'s (\ref{eq:2-068}) and (\ref{eq:2-069}), respectively,
  coincide with those proposed by Landau and Lifshitz,
  but
  those in the particle frame, given by Eq.'s (\ref{eq:2-082}) and (\ref{eq:2-083}),
  are slightly different from
  those given by Eckart and by Stewart.
  Our generic equation is also reduced to
  a novel relativistic hydrodynamic equation
  with the energy-momentum tensor and the particle current given by
  Eq.'s (\ref{eq:2-095}) and (\ref{eq:2-095}), respectively,
  where the bulk pressure (viscosity) term is absent.
  
  Furthermore,
  we have proved that
  the derived equation in a generic local rest frame
  has a stable equilibrium state
  owing to the positive definiteness of the inner product,
  although such a generic stability was suggested
  by a numerical calculation for some specific parameter sets
  in a previous paper by the present authors.
  It is worth emphasizing that
  all of our equations have a stable equilibrium state
  even in the particle frame in contrast to the Eckart and Stewart equations;
  note that such a drawback of the equation
  that the thermal equilibrium state can be unstable
  is taken over to some
  causal equations including the Israel-Stewart equation.
  
  In conformity with the above fact,
  we have proved that the Eckart equation can not be compatible
  with the underlying relativistic Boltzmann equation.
  This proof is based on the following significant observation
  that the orthogonality condition of the excited modes
  to the zero modes coincides with the ansatz's
  posed on the dissipative parts of the energy-momentum tensor and the particle current
  in the phenomenological equations.

  Thus,
  we have arrived at a sound starting point for
  attacking the last problem, i.e., constructing a causal relativistic hydrodynamic equation:
  We can expect that
  the present method applies
  to derive a causal equation on a sound basis,
  without recourse to any ansatz's
  as is done in other method such as the Maxwell-Grad moment method \cite{mic001,mic004,grad}.
  We have found \cite{next002}
  that a non-trivial extension of the P${}_0$ space in our formalism
  gives a causal equation,
  whose low-frequency limit is identical to the equation derived in this paper.
  This extension to derive causal equations will
  be reported in the forthcoming paper \cite{next002},
  where a correct moment method will be also proposed.
  
  We stress that
  all of the equations derived in this paper
  are consistent with the underlying kinetic equation,
  so the equations and also the method developed here may be useful
  for the analysis of the system
  where the proper dynamics describing the system changes
  from the hydrodynamic to kinetic regime,
  as in the system near the freeze-out region in the RHIC
  phenomenology \cite{dis005,nonaka06,connect}.
  
  Due to the absence of the problematic Burnett term
  in our relativistic equations,
  we expect that
  applying the RG method
  to the non-relativistic Boltzmann equation
  leads to
  the Navier-Stokes equation without the Burnett term,
  whose existence is known to be inevitable \cite{landau}
  when the Chapman-Enskog expansion method is applied.

  Finally,
  we emphasize that our method itself has
  a universal nature
  and can be applied to derive a slow dynamics
  from kinetic equations other than the simple Boltzmann equation.
  For example,
  it is successfully applicable
  to derive the dissipative hydrodynamic equation for the multi-component
  system in the energy frame
  from the multi-component Boltzmann equation \cite{mic001}.
  Furthermore,
  it would be interesting to apply the present method
  for extracting a Lorentz-covariant hydrodynamics
  of strongly interacting systems out of the equilibrium
  from the Kadanoff-Baym quantum-transport equation \cite{kadanoff}
  for the non-equilibrium many-body Green's function.
  
  \section*{
    Acknowledgements
  }
  We are grateful to K. Ohnishi for his collaboration in the early stage of this work.
  T.K. was partially supported by a
  Grant-in-Aid for Scientific Research by the Ministry of Education,
  Culture, Sports, Science and Technology (MEXT) of Japan (No.20540265, N0.23340067),
  by Yukawa International Program for Quark-Hadron Sciences, and by the
  Grant-in-Aid for the global COE program `` The Next Generation of
  Physics, Spun from Universality and Emergence '' from MEXT.
  
  \appendix
  
  \setcounter{equation}{0}
  \section{
    Most general form of phenomenological 
    relativistic dissipative hydrodynamic equations with
    dissipative internal energy $\delta e$, pressure $\delta p$,
    and density $\delta n$
  }
  The most general form of the dissipative parts of the energy-momentum tensor and
  particle current takes the following forms,
  \begin{eqnarray}
    \delta T^{\mu\nu} &=& \delta e \, u^\mu \, u^\nu - \delta p \, \Delta^{\mu\nu} +
    Q^\mu \, u^\nu + Q^\nu \, u^\mu + \Pi^{\mu\nu}, \\
    \delta N^{\mu} &=& \delta n \, u^\mu + \nu^\mu,
  \end{eqnarray}
  respectively.
  Note that we have made it explicit that
  the dissipations may cause corrections
  to the internal energy  $\delta e$, the pressure  $\delta p$ 
  and the particle-number density  $\delta n$,
  although only $\delta p$ is usually considered
  (as the bulk pressure term) in the literature.
  We remark that $\delta e$ and $\delta p$ are related to
  the dissipative part of the energy-momentum tensor, as follows,
  \begin{eqnarray}
    u_{\mu} \, \delta T^{\mu\nu} \, u_{\nu} &=& \delta e, \\
    \delta T^{\mu}_{\,\,\,\mu} &=& \delta e  - 3 \, \delta p.
  \end{eqnarray}
  
  In the present parametrization,
  $e+\delta e$, $p+\delta p$, and $n+\delta n$
  are the internal energy, pressure,
  and particle-number density in the dissipative system,
  with $e = e(T,\,\mu)$,   $p = p(T,\,\mu)$,   and $n = n(T,\,\mu)$
  being the corresponding quantities in the local equilibrium state
  with the temperature $T$ and the chemical potential $\mu$.
  In the tensor decomposition,
  \begin{eqnarray}
    Q^\mu &\equiv& \Delta^{\mu\nu} \, u^\rho \, T_{\nu\rho},\\
    \nu^\mu &\equiv& \Delta^{\mu\nu} \, N_\nu, \\
    \Pi^{\mu\nu}&\equiv& \Delta^{\mu\nu\rho\sigma} \, T_{\rho\sigma},
  \end{eqnarray}
  where $\Delta^{\mu\nu\rho\sigma}$ is
  the space-like, symmetric and traceless tensor
  defined in Eq.(\ref{eq:2-023}).
  
  Owing to the properties of the dissipative parts,
  $Q^\mu \, u_\mu = 0$, $\nu^\mu \, u_\mu = 0$,
  $\Pi^{\mu\nu} = \Pi^{\nu\mu}$,
  and $u_\mu \, \Pi^{\mu\nu} = \Pi^\mu_{\,\,\,\mu} = 0$,
  the total number of independent components
  of $Q^\mu$, $\nu^\mu$, and $\Pi^{\mu\nu}$ is eleven.
  Since $T^{\mu\nu}$ and $N^\mu$
  have fourteen components in total,
  $\delta e$, $\delta p$, and $\delta n$
  can not be independent of each other, but the number of independent
  component is one other than $T$ and $\mu$.
  We take $\delta p= \Pi$ as the independent component as a natural choice,
  then $\delta e$ and $\delta n$ may be expressed as
  $\delta e = f_e \, \Pi$ and $\delta n = f_n \, \Pi$,
  where $f_e$ and $f_n$ are functions of $T$ and $\mu$;
  $f_e = f_e(T,\,\mu)$ and $f_n = f_n(T,\,\mu)$.
  Here, we have assumed that the dissipative order of
  $\delta e$ and $\delta n$ is the same as
  that of $\delta p$ at most.
  We remark that in terms of $f_e$,
  \begin{eqnarray}
    u_{\mu} \, \delta T^{\mu\nu} \, u_{\nu} &=& f_e \, \Pi, \\
    \delta T^{\mu}_{\,\,\,\mu} &=&(f_e - 3) \, \Pi.
  \end{eqnarray}
  
  Now we shall show that
  the usual phenomenological derivation
  of the hydrodynamic equations
  allows the existence of $\delta e$ and $\delta n$, i.e.,
  finite values of $f_e$ and $f_n$,
  in the relativistic dissipative hydrodynamic equations.
  We emphasize that it can not be excluded on the general ground that 
  $f_e$ and $f_n$ may have finite values,
  although the phenomenological theory can not determine their 
  parametric forms as functions of $T$ and $\mu$.
  Although the first-order equation is considered in this article,
  the same argument equally applies to the second-order equations.
  
  Now the entropy current is given by
  \begin{eqnarray}
    \label{eq:A2-015}
    T \, S^\mu = p \, u^\mu + u_\nu \, T^{\mu\nu} -
    \mu \, N^\mu.
  \end{eqnarray}
  The second law of thermodynamics
  reads $\partial_\mu S^\mu \ge 0$.
  
  The divergence of $S^\mu$ is found to take the form
  \begin{eqnarray}
    \label{eq:A2-016}
    \partial_\mu S^\mu &=& \Pi \, \Bigg[
      f_e\,D\frac{1}{T} - \frac{1}{T} \, \nabla^\mu u_\mu
      - f_n\,D\frac{\mu}{T}
      \Bigg]
    + Q^\mu \, \Bigg[ \frac{1}{T} \, Du_\mu + \nabla_\mu
      \frac{1}{T} \Bigg]\nonumber \\
    & &   - \nu^\mu \,   \nabla_\mu \frac{\mu}{T}
    + \Pi^{\mu\nu} \, \frac{1}{T} \, \nabla_\mu u_\nu,
  \end{eqnarray}
  where
  $D \equiv u^\mu \, \partial_\mu$
  and  $\nabla^\mu \equiv \Delta^{\mu \nu} \, \partial_\nu$.
  Here, we have used
  the hydrodynamic equation and the first law of thermodynamics,
  \begin{eqnarray}
    D(p/T) + e \, D(1/T) - n \, D(\mu/T) = 0.
  \end{eqnarray}
  
  In the particle frame where $\nu^\mu = 0$,
  Eq.(\ref{eq:A2-016}) is reduced to
  \begin{eqnarray}
    \label{eq:A2-019}
    \partial_\mu S^\mu &=& \Pi \, \Bigg[
      f_e\,D\frac{1}{T} - \frac{1}{T} \, \nabla^\mu u_\mu
      - f_n\,D\frac{\mu}{T}
      \Bigg]
    + Q^\mu \, \Bigg[ \frac{1}{T} \, Du_\mu + \nabla_\mu \frac{1}{T} \Bigg]\nonumber \\
    & &  + \Pi^{\mu\nu} \, \frac{1}{T} \, \nabla_\mu u_\nu.
  \end{eqnarray}
  For assuring the second law of thermodynamics,
  one should make the right-hand side semi-positive definite.
  A natural choice is
  the following constitutive equations,
  \begin{eqnarray}
    \label{eq:A2-021}
    \Pi &=& \zeta \, T \, \Bigg[
      f_e\,D \frac{1}{T} - \frac{1}{T} \, \nabla^\mu u_\mu
      - f_n\,D\frac{\mu}{T} \Bigg],\\
    \label{eq:A2-022}
    Q^\mu &=& - \lambda \, T^2 \, \Bigg[ \frac{1}{T} \, Du^\mu +
      \nabla^\mu \frac{1}{T} \Bigg],\\
    \label{eq:A2-023}
    \Pi^{\mu\nu} &=& 2 \, \eta \, \Delta^{\mu\nu\rho\sigma} \,
    \nabla_\rho u_\sigma,
  \end{eqnarray}
  with $\zeta$, $\lambda$, and $\eta$
  being the bulk viscosity, heat conductivity, and shear viscosity, respectively.
  Indeed,
  the above constitutive equations lead to
  \begin{eqnarray}
    \label{eq:A2-020}
    \partial_\mu S^\mu
    = \frac{\Pi^2}{\zeta T}
    - \frac{Q^\mu Q_\mu}{\lambda T^2}
    + \frac{\Pi^{\mu\nu}\Pi_{\mu\nu}}{2\eta T} \ge 0.
  \end{eqnarray}

  Thus, one sees that
  the relativistic dissipative hydrodynamic equations
  with  finite $f_e$ and $f_n$, or equivalently
  with finite $\delta e$ and $\delta n$,
  is compatible with the second law of thermodynamics.
  With a special choice, $f_e = f_n = 0$,
  Eq.'s (\ref{eq:A2-021})-(\ref{eq:A2-023}) lead
  to the constitutive equations proposed by Eckart
  that are commonly used.
  
  In the case of the energy frame where $Q^\mu = 0$,
  a similar argument leads to Eq.'s (\ref{eq:A2-021}), (\ref{eq:A2-023}), and
  \begin{eqnarray}
    \label{eq:A2-024}
    \nu^\mu = \lambda \, \hat{h}^{-2} \,
    \nabla^\mu \frac{\mu}{T},
  \end{eqnarray}
  with $\hat{h} = (e + p)/n\,T$ being the reduced enthalpy per particle.
  We remark that
  these equations are reduced to
  the constitutive equations by Landau and Lifshitz
  only when one can set $f_e = f_n = 0$.
  
  The phenomenological theory can not proceed further
  to specify $f_e$ and $f_n$;
  the values of $f_e$ and $f_n$ can be determined only from a microscopic theory.
  In the present work, we show that the microscopic theory
  leads to
  $f_e = f_n = 0$ in the energy frame,
  while
  $f_e = 3$ and $f_n = 0$ in the particle frame,
  implying that
  $\delta T^{\mu}_{\,\,\,\mu} = 0$ but
  $u_{\mu} \, \delta T^{\mu\nu} \, u_{\nu} = 3 \, \Pi \neq 0$.
  
  \setcounter{equation}{0}
  \section{
    Brief account of the renormalization-group method with an example
  }
  In this Appendix,
  we first show how
  the renormalization-group (RG) method \cite{rgm001,env001,env002,env005,qm,env006} works
  as a reduction theory of the dynamics,
  adopting van der Pol equation with a limit cycle,
  as an example.
  Then,
  we briefly present a foundation of the RG method.
  A detailed account of the RG method including its foundation 
  may be seen in Ref.'s \citen{env001,env002,qm,env006}
  or review articles \citen{env008,Butsuri}.
  
  \subsection{
    RG method applied to van der Pol equation
  }
  Let us take the van der Pol equation
  which admits a limit cycle:
  \begin{eqnarray}
    \ddot{x} + x = \eps \, (1 - x^2) \, \dot{x},
    \label{eq:van-der-pol}
  \end{eqnarray}
  where $\eps$ is supposed to be small.
  
  Let us solve the same problem by the RG method.
  Let $\tilde{x}(t;\,t_0)$ be a local solution
  around $t\sim \forall t_0$, and
  represent it as a perturbation series;
  $\tilde{x}(t;\,t_0) = \tilde{x}_0(t;\,t_0)
  +\eps \, \tilde{x}_1(t;\,t_0) + \eps^2 \, \tilde{x}_2(t;\,t_0) + \cdots$.
  In the RG method,
  the initial value $W(t_0)$ matters:
  We suppose that an exact solution is given by $x(t)$
  and the initial value of $\tilde{x}(t;\,t_0)$ at $t=t_0$ is set up to be $x(t_0)$;
  i.e., $W(t_0)\equiv \tilde{x}(t_0;\,t_0) = x(t_0)$.
  The initial value as the exact solution should be also expanded as
  $W(t_0) = W_0(t_0) + \eps \, W_1(t_0) + \eps^2 \, W_2(t_0) + \cdots$.
  The RG method is actually the method
  to obtain the initial value $W(t)$
  as an exact solution or approximate solution
  valid in a global domain asymptotically.
  
  The zeroth-order equation reads
  \begin{eqnarray}
    {\cal L}\tilde{x}_0 \equiv \left[ \frac{d^2}{dt^2} + 1 \right]
    \tilde{x}_0 = 0,
  \end{eqnarray}
  the solution to which can be written as
  \begin{eqnarray}
    \tilde{x}_0(t;\,t_0) = A(t_0) \, \cos(t + \theta(t_0)).
  \end{eqnarray}
  Here,
  we have made it explicit that
  the integral constants $A$ and $\theta$ may depend on the
  initial time $t_0$.
  The equation for $\tilde{x}_1$ reads
  \begin{eqnarray}
    {\cal L}\tilde{x}_1 =
    - A \, \left( 1 - \frac{A^2}{4} \right) \, \sin\phi(t)
    + \frac{A^3}{4} \, \sin 3\,\phi(t),
  \end{eqnarray}
  with $\phi(t) = t + \theta_0(t_0)$.
  Notice that the first term in the right-hand side is
  a zero mode of the linear operator ${\cal L}$ appearing in the
  left-hand side.
  Thus, the special solution to this equation
  necessarily contains a secular term
  which is given by $t$ times a zero mode of ${\cal L}$.
  Since we have supposed that
  the initial value at $t=t_0$ is on an exact solution,
  the corrections from the zeroth-order solution
  should be as small as possible.
  This condition is realized by setting the secular terms
  appearing in the higher orders vanish at $t=t_0$,
  which is possible because we can add freely zero mode solutions to
  a special solution.
  Thus,
  the first-order solution is uniquely written as
  \begin{eqnarray}
    \tilde{x}_1(t;t_0) = (t-t_0) \, \frac{A}{2} \, \left(1 -
    \frac{A^2}{4}\right) \, \sin \phi(t)
    - \frac{A^3}{32} \, \sin 3\,\phi(t).
    \label{eq:van-der-Pol-RG-1sr}
  \end{eqnarray}
  Notice that the secular term surely vanishes at $t=t_0$
  in Eq.(\ref{eq:van-der-Pol-RG-1sr}),
  implying that its initial value at $t=t_0$ reads
  \begin{eqnarray}
    \tilde{x}_1(t_0;t_0) = - \frac{A^3(t_0)}{32} \, \sin 3\,\phi(t_0).
    \label{eq:van-der-pol-rg-init-1}
  \end{eqnarray}
  
  The perturbative solution up to this order
  reads $\tilde{x} = \tilde{x}_0 + \eps \, \tilde{x}_1$,
  which becomes, however, invalid
  when $\vert t-t_0 \vert \rightarrow $ large,
  because of the secular term.
  
  Now notice that the function $\tilde{x}(t; t_0)$ corresponds to
  a curve drawn in the $(t,\,x)$ plane for each $t_0$;
  in other words,
  we have a family of curves represented by $\tilde{x}(t;t_0)$
  in the $(t,\,x)$ plane,
  a member of which is parametrized by $t_0$.
  An important observation is that
  each curve is close to the exact solution in the neighborhood of
  $t = t_0$.
  Thus,
  an idea is that the envelope curve of the family of curves
  should give a global solution.
  The envelope curve can be constructed by solving
  the following equation
  \begin{eqnarray}
    \frac{\mathrm{d}\tilde{x}}{\mathrm{d}t_0} \Big \vert_{t_0=t} = 0,
    \label{eq:van-der-pol-env}
  \end{eqnarray}
  which leads to the following equations for $A(t)$ and $\phi(t)$,
  \begin{eqnarray}
    \dot{A} &=& \eps \, \frac{A}{2} \, \left( 1-\frac{A^2}{4} \right),
    \label{eq:amplitude}\\
    \dot{\phi} &=& 1.
    \label{eq:phase}
  \end{eqnarray}
  These equation are readily solved,
  and one sees that
  as $t \rightarrow \infty$,
  $A(t) \rightarrow 2$ asymptotically,
  meaning the existence of a limit cycle with a radius $2$.
  
  The resultant envelope function as a global solution
  is given by
  \begin{eqnarray}
    x_{\mathrm{E}}(t) \equiv \tilde{x}(t;\, t) = W(t)
    = A(t)\,\cos(t+\theta_0) - \eps \,\frac{A^3(t)}{32} \, \sin (3\,t
    + 3\,\theta_0),
  \end{eqnarray}
  with $A(t)$ being the solution of Eq.(\ref{eq:amplitude}).
  Thus,
  we have succeeded in not only obtaining
  the asymptotic solution as whole but also extracting
  the slow variables $A(t)$ and $\phi(t)$
  explicitly and their governing equations.
  
  One now sees that
  when there exist zero modes of the unperturbed operator,
  the higher-order corrections may cause secular terms,
  which are renormalized into
  the integral constants in the zeroth-order solution by the RG/envelope equation
  (\ref{eq:van-der-pol-env}),
  and thereby the would-be integral constants
  are lifted to dynamical but slow variables.
  
  In the derivation of hydrodynamic equations from the kinetic equation,
  the would-be integral constants corresponding to $A$ and $\phi$ are
  the temperature $T$,
  the chemical potential $\mu$,
  and the flow velocity $u^{\mu}$ ($u^{\mu} \, u_{\mu} = 1$)
  characterizing the local equilibrium state,
  which are to be lifted to the slow dynamical variables
  and their governing equations are identically the hydrodynamic equation.
  
  \subsection{
    Foundation of RG method
  }
  Now we present a foundation to the RG method
  using a Wilsonian equation \cite{wilson}
  or flow equation by Wegener \cite{weg}.
  
  Let us take the following $n$-dimensional equation;
  \begin{eqnarray}
    \label{eq:B3-1}
    \frac{\mathrm{d} \bfX}{\mathrm{d}t} = \bfF(\bfX, t),
  \end{eqnarray}
  where $n$ may be infinity.
  Let $\bfX(t) = \bfW(t)$ be an yet unknown exact solution to Eq.(\ref{eq:B3-1}),
  and we try to solve the equation
  with the initial condition at $t=\forall t_0$;
  \begin{eqnarray}
    \label{init0}
    \bfX(t=t_0) = \bfW(t_0).
  \end{eqnarray}
  Then, the solution may be written as $\bfX(t; t_0, \bfW(t_0))$.
  
  Now the basis of the RG method lies in the fact that
  $\bfW(t_0)$ can be determined on the basis of a simple fact of differential equations.
  We notice that
  when the  initial point is shifted to $t_0'$,
  the resultant solution should be the same as long as $\bfW(t)$ is an exact solution, i.e.,
  \begin{eqnarray}
    \bfX(t; t_0, \bfW(t_0)) = \bfX(t; t_0', \bfW(t_0')).
  \end{eqnarray}
  Taking the limit $t_0' \rightarrow t_0$,
  we have
  \begin{eqnarray}
    \label{rg00}
    \frac{\mathrm{d} \bfX}{\mathrm{d}t_0} =\frac{\d \bfX}{\d t_0}+
    \frac{\d \bfX}{\d \bfW}\frac{\mathrm{d} \bfW}{\mathrm{d}t_0} = {\bf 0}.
  \end{eqnarray}
  This equation gives an evolution equation or
  the flow equation of the initial value $\bfW(t_0)$.
  This equation has the same form as and corresponds to the non-perturbative RG equations
  (flow equations) by Wilson \cite{wilson},
  Wegner-Houghton \cite{weg},
  and so on, in quantum field theory and statistical physics:
  The 'initial time' $t_0$ corresponds to 
  (the logarithm of) the renormalization point.
  We emphasize that the equation (\ref{rg00}) is exact;
  we have no recourse to
  any perturbation theory so far.
  
  The problem is to construct the seed of the RG equation (\ref{rg00}), i.e.,
  $\bfX(t; t_0, \bfW(t_0))$.
  For that,
  let us take the perturbation theory.
  In this case, $\bfX (t;t_0, \bfW(t_0))$ and $\bfX (t;t'_0, \bfW(t'_0))$
  may  be valid only for $t\sim t_0$ and $t\sim t'_0$,
  which condition is naturally satisfied
  when $t_0<t<t'_0$ (or $t'_0<t<t_0$) because
  the limit $t'_0 \rightarrow t_0$ is taken eventually.
  Thus,
  when a perturbative expansion is employed for constructing
  $\bfX (t; t_0, \bfW (t_0))$,
  it is necessary to demand
  \begin{eqnarray} 
    \label{rg0}
    \frac{\mathrm{d} \bfX}{\mathrm{d}t_0} \biggl\vert _{t_0=t}
    = \frac{\d \bfX}{\d t_0}
    \biggl\vert _{t_0=t} +\frac{\d \bfX}{\d \bfW}\frac{\mathrm{d} \bfW}{\mathrm{d}t_0}
    \biggl\vert _{t_0=t} ={\bf 0},
  \end{eqnarray}
  which automatically implies the condition that $t_0=t$.
  
  We have already suggested an interpretation
  that this equation can be
  identified as a condition to construct an envelope of the curves represented
  by the unperturbed solutions with different initial times $t_0$'s \cite{env001}:
  When $t_0$ is varied,
  $\bfX(t; t_0, \bfW(t_0))$
  gives a family of curves with $t_0$ being a parameter characterizing curves.
  Then, Eq.(\ref{rg0}) is a condition  to construct the envelope of
  the family of curves which are valid only locally around $t\sim t_0$.
  The envelope is given by $\bfX (t;t_0=t)=\bfW (t)$, i.e, the initial value.
  
  Now we shall show that
  $\bfX (t;t_0=t)=\bfW(t)$ satisfies the original equation
  (\ref{eq:B3-1}) in a global domain up to the order with which
  $\bfX(t; t_0)$ satisfies around $t\sim t_0$. 
  Let $\bfX(t;t_0)$ is an approximate solution to
  Eq.(\ref{eq:B3-1}) around $t\sim t_0$;
  \begin{eqnarray}
    \frac{\mathrm{d} \bfX(t; t_0)}{\mathrm{d}t} \simeq \bfF(\bfX(t; t_0), t).
  \end{eqnarray}
  Then, we have
  \begin{eqnarray}
    \frac{\mathrm{d}\bfW(t)}{\mathrm{d} t} &=& \frac{\d \bfX(t;t_0)}{\d t}\biggl\vert_{t_0=t}
    +\frac{\d \bfX(t;t_0)}{\d t_0}\biggl\vert_{t_0=t}\nonumber\\
    &=& \frac{\d \bfX(t;t_0)}{\d t}\biggl\vert_{t_0=t}\nonumber \\
    &\simeq& \bfF(\bfX(t; t_0), \, t)\vert_{t_0=t}, \nonumber \\
    &=& \bfF(\bfW(t), t),
  \end{eqnarray}
  on account of Eq.(\ref{rg0}).
  This proves the above statement.
  
  \setcounter{equation}{0}
  \section{
    Derivation of first-excited modes with
    a generic macroscopic-frame Vector
  }
  \label{sec:8}
  In this Appendix,
  we explicitly derive
  the first-excited modes $\varphi^{\mu\alpha}_{1p}$
  in the generic local rest frame with
  the macroscopic-frame vector being specified by Eq.(\ref{eq:2-006}),
  i.e.,
  $\Ren{a}^\mu_p = \theta^\mu_p =
  [((p\cdot u)\,\cos\theta + m\,\sin\theta)/(p\cdot u)] \, u^\mu$.
  
  For convenience,
  we introduce
  a dimensionless quantity \cite{mic001} dependent on $z = m/T$,
  \begin{eqnarray}
    \label{eq:6-001}
    a_\ell &\equiv& \frac{1}{n \, T^{\ell-1}} \, \sum_p \, \frac{1}{p^0} \,
    f^\mathrm{eq}_p \, (p\cdot u)^\ell\nonumber\\
    &=& \frac{1}{z^2\,K_2(z)} \, \int_z^\infty\!\!\mathrm{d}\tau \,\,\,
    (\tau^2 - z^2)^{1/2} \, \tau^\ell \, \mathrm{e}^{-\tau}.
  \end{eqnarray}
  This quantity for $\ell = 0, 1, 2, 3, 4, 5$ reads
  \begin{eqnarray}
    \label{eq:6-002}
    a_0 &=& z^{-2} \, (\hat{h} - 4),\\
    \label{eq:6-003}
    a_1 &=& 1,\\
    \label{eq:6-004}
    a_2 &=& \hat{h} - 1,\\
    \label{eq:6-005}
    a_3 &=& 3 \, \hat{h} + z^2,\\
    \label{eq:6-006}
    a_4 &=& (15 + z^2) \, \hat{h} + 2\,z^2,\\
    \label{eq:6-007}
    a_5 &=& 6\,(15 + z^2) \, \hat{h} + z^2 \, (15 + z^2),
  \end{eqnarray}
  respectively.
  Here,  $\hat{h}$ is defined in Eq.(\ref{eq:2-021}) and
  $K_\ell(z)$ denotes the modified Bessel function,
  \begin{eqnarray}
    \label{eq:6-008}
    K_\ell(z) = \frac{2^\ell\,\ell!}{(2\,\ell)!}\,z^{-\ell}\,
    \int_z^\infty\!\!\mathrm{d}\tau \,\,\,
    (\tau^2 - z^2)^{\ell - 1/2} \, \mathrm{e}^{-\tau},
  \end{eqnarray}
  which satisfies the recurrence relation
  \begin{eqnarray}
    \label{eq:6-009}
    K_{\ell+1}(z) = K_{\ell-1}(z) + \frac{2\,\ell}{z} \, K_\ell(z).
  \end{eqnarray}
  
  First,
  the setting of $\Ren{a}^\mu_p = \theta^\mu_p$ leads us to
  the metric matrix $\eta^{\alpha\beta}_0$ given by
  \begin{eqnarray}
    \label{eq:6-010}
    \eta_0^{\mu\nu} &=& n \, T^2 \, \Bigg[(a_3 \, \cos\theta + z \, a_2 \,
      \sin\theta) \, u^\mu \, u^\nu\nonumber\\
      &&{}+ \Big((z^2 \, a_1 - a_3) \,
      \cos\theta + z \, (z^2 \, a_0 - a_2) \,
      \sin\theta \Big) \, \frac{1}{3} \, \Delta^{\mu\nu}\Bigg],\\
    \label{eq:6-011}
    \eta_0^{\mu 4} = \eta_0^{4\mu} &=& n \, T^2 \, z \, (a_2 \,
    \cos\theta + z \, a_1 \,
    \sin\theta) \, u^\mu,\\
    \label{eq:6-012}
    \eta_0^{44} &=& n \, T^2 \, z^2 \, (a_1 \,
    \cos\theta + z \, a_0 \,
    \sin\theta).
  \end{eqnarray}
  By a tedious but straightforward manipulation,
  we have
  \begin{eqnarray}
    \label{eq:6-013}
    \eta^{-1}_{0\mu\nu} &=& (n \, T^2)^{-1} \, \Big( A(\theta) \, u_\mu \, u_\nu +
    B(\theta) \, \Delta_{\mu\nu} \Big),\\
    \label{eq:6-014}
    \eta^{-1}_{0\mu 4} = \eta^{-1}_{04\mu} &=& (n \, T^2)^{-1} \, C(\theta) \, u_\mu,\\
    \label{eq:6-015}
    \eta^{-1}_{044} &=& (n \, T^2)^{-1} \, D(\theta),
  \end{eqnarray}
  where
  \begin{eqnarray}
    \label{eq:6-016}
    A(\theta) &\equiv& \frac{a_1 \, \cos\theta + a_0 \, z \, \sin\theta}
    {(a_3 \, a_1 - a^2_2) \, \cos^2\theta +
      (a_3 \, a_0 - a_2 \, a_1) \, z \, \sin\theta \, \cos\theta +
      (a_2 \, a_0 - a^2_1) \, z^2 \, \sin^2\theta},\nonumber\\\\
    \label{eq:6-017}
    B(\theta) &\equiv& \frac{3}{(z^2 \, a_1 - a_3) \, \cos\theta + (z^2
      \, a_0 - a_2) \, z \, \sin\theta},\\
    \label{eq:6-018}
    C(\theta) &\equiv& - \frac{1}{z} \, \frac{a_2 \, \cos\theta + a_1 \, z \, \sin\theta}
    {(a_3 \, a_1 - a^2_2) \, \cos^2\theta +
      (a_3 \, a_0 - a_2 \, a_1) \, z \, \sin\theta \, \cos\theta +
      (a_2 \, a_0 - a^2_1) \, z^2 \, \sin^2\theta},\nonumber\\\\
    \label{eq:6-019}
    D(\theta) &\equiv& \frac{1}{z^2} \, \frac{a_3 \, \cos\theta + a_2 \, z \, \sin\theta}
    {(a_3 \, a_1 - a^2_2) \, \cos^2\theta +
      (a_3 \, a_0 - a_2 \, a_1) \, z \, \sin\theta \, \cos\theta +
      (a_2 \, a_0 - a^2_1) \, z^2 \, \sin^2\theta}.\nonumber\\
  \end{eqnarray}
  
  Next, the inner product
  $\langle\, \varphi^\alpha_0 \,,\, \tilde{\varphi}^{\mu\beta}_1 \,\rangle$ is evaluated to be
  \begin{eqnarray}
    \label{eq:6-020}
    \langle\, \varphi^a_0 \,,\, \tilde{\varphi}^{\mu b}_1 \,\rangle &=&
    n\,T^2 \, \Big(
    a_3 \, u^a \, u^\mu \, u^b +
    (z^2\,a_1 - a_3)\,
    \frac{1}{3}\,
    (u^a \, \Delta^{\mu b} + u^\mu \, \Delta^{ba} + u^b \, \Delta^{a\mu})\Big)
    ,\nonumber\\\\
    \label{eq:6-021}
    \langle\, \varphi^a_0 \,,\, \tilde{\varphi}^{\mu 4}_1 \,\rangle
    &=&
    n\,T^2 \, \Big(z\,a_2\,u^a\,u^\mu + z\,(z^2\,a_0 - a_2)\,\frac{1}{3}\,\Delta^{a\mu}\Big)
    ,\\
    \label{eq:6-022}
    \langle\, \varphi^4_0 \,,\, \tilde{\varphi}^{\mu b}_1 \,\rangle
    &=&
    n\,T^2 \, \Big(z\,a_2\,u^\mu\,u^b + z\,(z^2\,a_0 -
    a_2)\,\frac{1}{3}\,\Delta^{\mu b}\Big)
    ,\\
    \label{eq:6-023}
    \langle\, \varphi^4_0 \,,\, \tilde{\varphi}^{\mu 4}_1 \,\rangle
    &=&
    n\,T^2\,z^2\,a_1\,u^\mu.
  \end{eqnarray}
  
  Finally,
  combining $\eta^{-1}_{0\alpha\beta}$ in Eq.'s (\ref{eq:6-013})-(\ref{eq:6-015}) and
  $\langle\, \varphi^\alpha_0 \,,\, \tilde{\varphi}^{\mu\beta}_1 \,\rangle$
  in Eq.'s (\ref{eq:6-020})-(\ref{eq:6-023}) together with
  \begin{eqnarray}
    \label{eq:6-024}
    \varphi^{\mu\alpha}_{1p}
    = \big[ Q_0 \, \tilde{\varphi}^{\mu\alpha}_1 \big]_p
    &=& \tilde{\varphi}^{\mu\alpha}_{1p}
    - \big[ P_0 \, \tilde{\varphi}^{\mu\alpha}_1 \big]_p\nonumber\\
    &=& \frac{1}{p\cdot\theta_p} \, \Big( p^\mu\,\varphi^\alpha_{0p}
    - (p\cdot\theta_p) \,
    \varphi^\beta_{0p}\,\eta^{-1}_{0\beta\gamma}\,
    \langle\, \varphi^\gamma_0\,,\,\tilde{\varphi}^{\mu\alpha}_1 \,\rangle
    \Big),
  \end{eqnarray}
  we have
  \begin{eqnarray}
    \label{eq:6-025}
    \varphi^{\mu\alpha}_{1p} =\left\{
    \begin{array}{ll}
      \displaystyle{
        \frac{1}{p\cdot\theta_p} \, \Bigg[
  	\Pi_p\,\Big(
  	u^\mu\,u^\nu\,Y_1(\theta) - \Delta^{\mu\nu}\,Y_2(\theta)
  	\Big)
      }
      &
      \displaystyle{
      }\\[2mm]
      \displaystyle{
        \hspace{2cm}{}+ (J^\mu_p\,u^\nu + J^\nu_p\,u^\mu)\,Y_3(\theta)
        + \pi^{\mu\nu}_p
        \Bigg]
      }
      &
      \displaystyle{
        \mathrm{for}\,\,\,\alpha=\mu,
      }\\[2mm]
      \displaystyle{
        \frac{1}{p\cdot\theta_p} \, \Bigg[
  	\Pi_p\,u^\mu\,Z_1(\theta) + J^\mu_p \,Z_2(\theta)
  	\Bigg]
      }
      &
      \displaystyle{
        \mathrm{for}\,\,\,\alpha=4.
      }
    \end{array}
    \right.
  \end{eqnarray}
  Here, we have introduced the following quantities
  \begin{eqnarray}
    \label{eq:6-026}
    \Pi_p &\equiv& \frac{
      (a_2\,a_0 - a^2_1) \, z^2 \, (p\cdot u)^2
      - (a_3\,a_0 - a_2\,a_1) \, z \, m \, (p\cdot u) + (a_3\,a_1 - a^2_2)
      \, m^2
    }{
      - 3\,(a_3\,a_1 - a^2_2)
    },\nonumber\\\\
    \label{eq:6-027}
    J^\mu_p &\equiv& \frac{
      \Delta^{\mu\nu}\,p_\nu\,
      \Big[ (z^2\,a_0-a_2) \, z \, (p\cdot u) - (z^2 \, a_1 - a_3) \, m
        \Big]
    }{
      - (z^2\,a_0 - a_2)\,z
    }
    ,\\
    \label{eq:6-028}
    \pi^{\mu\nu}_p &\equiv& \Delta^{\mu\nu\rho\sigma} \,
    p_\rho \, p_\sigma,\\
    \label{eq:6-029}
    Y_1(\theta) &\equiv& \frac{
      -3\,(a_3\,a_1-a^2_2)\,\sin^2\theta
    }{
      (a_3\,a_1-a^2_2)\,\cos^2\theta + (a_3\,a_0 -
      a_2\,a_1)\,z\,\sin\theta\,\cos\theta +
      (a_2\,a_0-a^2_1)\,z^2\,\sin^2\theta
    },\nonumber\\\\
    \label{eq:6-030}
    Y_2(\theta) &\equiv& \frac{
      -(a_3\,a_1-a^2_2)\,(\cos^2\theta - \sin^2\theta)
    }{
      (a_3\,a_1-a^2_2)\,\cos^2\theta + (a_3\,a_0 -
      a_2\,a_1)\,z\,\sin\theta\,\cos\theta +
      (a_2\,a_0-a^2_1)\,z^2\,\sin^2\theta
    },\nonumber\\\\
    \label{eq:6-031}
    Y_3(\theta) &\equiv& \frac{
      - (z^2\,a_0-a_2) \, z \, \sin\theta
    }{
      (z^2\,a_1 - a_3)\,\cos\theta + (z^2\,a_0 - a_2)\,z\,\sin\theta
    },\\
    \label{eq:6-032}
    Z_1(\theta) &\equiv& \frac{
      3\,(a_3\,a_1-a^2_2)\,\cos\theta\,\sin\theta
    }{
      (a_3\,a_1-a^2_2)\,\cos^2\theta + (a_3\,a_0 -
      a_2\,a_1)\,z\,\sin\theta\,\cos\theta +
      (a_2\,a_0-a^2_1)\,z^2\,\sin^2\theta
    },\nonumber\\\\
    \label{eq:6-033}
    Z_2(\theta) &\equiv& \frac{
      (z^2\,a_0-a_2) \, z \, \cos\theta
    }{
      (z^2\,a_1 - a_3)\,\cos\theta + (z^2\,a_0 - a_2)\,z\,\sin\theta
    }.
  \end{eqnarray}
  Substituting $a_0$, $a_1$, $a_2$, and $a_3$ in Eq.'s (\ref{eq:6-002})-(\ref{eq:6-007})
  into the above equations,
  we arrive at the explicit representation of $\varphi^{\mu\alpha}_{1p}$
  shown in Eq.(\ref{eq:2-011}).
  

\end{document}